\documentclass[english,notitlepage,nofootinbib,superscriptaddress]{revtex4-1}
\usepackage[T1]{fontenc}
\usepackage[latin9]{inputenc}
\usepackage{color}
\usepackage{amsmath}
\usepackage{graphicx}
\usepackage[normalem]{ulem}
\usepackage{soul,xcolor}

\setstcolor{red}
\newcommand\del{\bgroup\markoverwith
{\textcolor{red}{\rule[0.5ex]{2pt}{1.2pt}}}\ULon}

\makeatletter

\providecommand{\tabularnewline}{\\}

\makeatother

\usepackage{babel}
\begin{document}


\title{Vacuum structure of the left-right symmetric model}



\author{P.S. Bhupal Dev}
\email{bdev@physics.wustl.edu}
\affiliation{Department of Physics and McDonnell Center for the Space Sciences, Washington University, St.~Louis, MO 63130, USA}
\author{Rabindra N. Mohapatra}
\email{rmohapat@physics.umd.edu}
\affiliation{Maryland Center for Fundamental Physics and Department of Physics, University of Maryland, College Park, MD 20742, USA}
\author{Werner Rodejohann}
\email{werner.rodejohann@mpi-hd.mpg.de}
\affiliation{Max-Planck-Institut f\"ur Kernphysik, Postfach
103980, D-69029 Heidelberg, Germany}
\author{Xun-Jie Xu}
\email{xunjie@mpi-hd.mpg.de}
\affiliation{Max-Planck-Institut f\"ur Kernphysik, Postfach
103980, D-69029 Heidelberg, Germany}

\date{\today}

\begin{abstract}
\noindent 
The left-right symmetric model (LRSM), originally proposed to explain parity violation in low energy processes, has since emerged as an attractive framework for light neutrino masses via the seesaw mechanism. The scalar sector of the minimal LRSM consists of an $SU(2)$ bi-doublet, as well as left- and right-handed weak isospin triplets, thus making the corresponding vacuum structure much more complicated than that of the Standard Model. In particular, the desired ground state of the Higgs potential should be a charge conserving, and preferably global, minimum with parity violation at low scales. We show that this is not a generic feature of the LRSM potential and happens only for a small fraction of the parameter space of the potential. 
We also analytically study the potential for some simplified cases and obtain useful conditions (though not necessary) to achieve successful symmetry breaking.
We then carry out a detailed statistical analysis of the minima of the Higgs potential using numerical minimization and find that for a large fraction of the parameter space, the potential does not have a good vacuum. Imposing the analytically obtained conditions, we can readily find a small part of the parameter space with good vacua.
Consequences for some scalar masses are also discussed. 

\end{abstract}
\maketitle

\section{\label{sec:intro}Introduction}
\noindent The discovery of neutrino masses is a sure sign of new physics beyond the Standard Model (SM). A simple paradigm for neutrino masses is the seesaw mechanism~\cite{seesaw1, seesaw2, seesaw3, seesaw4, seesaw5} which introduces  right-handed neutrinos (RHN)  with  heavy Majorana masses. Two questions then arise: (i) what is the seesaw scale or the mass of the RHNs? and (ii) what is the ultraviolet (UV)-complete theory that leads naturally to the basic ingredients inherent in the seesaw mechanism i.e.\ RHNs and a $B-L$ symmetry whose breaking gives rise to their Majorana masses? Two classes of theories with this property are: (i) the $SO(10)$ model whose basic spinor representation contains the RHN and which contains a group generator that is the $B-L$ local symmetry~\cite{Fritzsch:1974nn}, and (ii) the left-right symmetric model (LRSM)~\cite{LR1, LR2, LR3}, which is the simplest extension of the SM that contains three RHNs to cancel the gauge anomalies and $B-L$ symmetry as a natural symmetry~\cite{Mohapatra:1980qe}.

In this paper, we focus on symmetry  breaking aspects of the minimal LRSM and carry out an analysis of its vacuum structure. The general procedure to investigate this is to write down the Higgs potential involving the various scalar multiplets of the LR gauge group $SU(2)_L\times SU(2)_R\times U(1)_{B-L}$ and look for the minimum of the potential that breaks the gauge symmetry down to $U(1)_{\rm em}$. Morever, the theory should be parity violating at low scale and generate naturally small neutrino masses. The
detailed analysis of the non-supersymmetric LRSM Higgs potential and its minima have been discussed in many works~\cite{Branco:1985ng, Basecq:1985sx, Gunion:1989in, Deshpande:1990ip,Choi:1992wb,Brahmachari:1994ts,Barenboim:2001vu, Kiers:2005gh, Chakrabortty:2013zja, Chakrabortty:2013mha, Mohapatra:2014qva, Mondal:2015fja, Dev:2016dja, Maiezza:2016bzp, Chakrabortty:2016wkl, Maiezza:2016ybz}   (supersymmetric LRSM Higgs sectors have been studied in Refs.~\cite{Babu:2014vba, Basso:2015pka}). It is known that for certain ranges of the parameters (e.g.\ negative scalar mass squares and positive values for scalar couplings), a desired (good) vacuum is obtained. However, if all the couplings are chosen randomly to start with, it is not clear how often one gets a good vacuum. Secondly, it is not known whether those minima obtained above are global minima of the potential or are simply the local ones. Third, the boundedness-from-below of the potential has been used as the necessary and sufficient condition for vacuum stability~\cite{Chakrabortty:2013zja, Chakrabortty:2013mha}, but as we show in this paper, a bounded-from-below potential is necessary for a good vacuum, but not sufficient.   

We use the gauge freedom of the theory to give simple criteria that are to be fulfilled in order to end up in a good vacuum (i.e.\ charge conserving and parity violating at low scales). 
Keeping arbitrary values for the parameters of the Higgs potential we furthermore check for what fraction of the parameter space a global minimum with desired properties is obtained.  
Once a range of the parameters is determined where a global minimum occurs, one can then use them to find the scalar spectrum corresponding to that choice, which is a potential test of the model. We do not carry out an exhaustive analysis of the scalar masses but rather give some simple examples at viable minima of the model.

The paper is organized as follows. In Sec.\ \ref{sec:model}, we review the details of the model and its scalar sector. In Sec.~\ref{sec:basic},
we write down the full scalar potential and give criteria for obtaining the good vacua of the model. In Sec.~\ref{sec:analytical}
we present analytical studies focused on some simplified cases, followed by numerical studies to scan the whole parameter space in Sec.~\ref{sec:numerical}. 
Finally we summarize and conclude in Sec.~\ref{sec:Conclusion}. Some technical details are delegated to the appendix.

\section{\label{sec:model}Model details}
\noindent
The LRSM~\cite{LR1, LR2, LR3} extends the SM gauge group ${\cal G}_{\rm SM}\equiv SU(3)_c\times SU(2)_L\times U(1)_Y$ to ${\cal G}_{\rm LR}\equiv SU(3)_c\times SU(2)_L\times SU(2)_R\times U(1)_{B-L}$. The quarks and leptons are assigned to the following irreducible representations of ${\cal G}_{\rm LR}$:
\begin{align}
& Q_{L,i} \ = \ \left(\begin{array}{c}u_L\\d_L \end{array}\right)_i : \: \left({ \bf 3},\ {\bf 2},\ {\bf 1},\ \frac{1}{3}\right), \qquad \qquad
Q_{R,i} \ = \ \left(\begin{array}{c}u_R\\d_R \end{array}\right)_i : \: \left({ \bf 3},\ {\bf 1},\ {\bf 2},\ \frac{1}{3}\right), \\
& \psi_{L,i} \ = \  \left(\begin{array}{c}\nu_L \\ e_L \end{array}\right)_i : \: \left({ \bf 1},\ {\bf 2},\ {\bf 1},\ -1 \right), \qquad \qquad
\psi_{R,i} \ = \ \left(\begin{array}{c} N_R \\ e_R \end{array}\right)_i : \: \left({ \bf 1},\ {\bf 1},\ {\bf 2},\ -1 \right),
\label{lrSM}
\end{align}
where $i=1,2,3$ represents the family index, and the subscripts $L,R$ denote  the left- and right-handed chiral projection operators $P_{L,R} = (1\mp \gamma_5)/2$, respectively. The $B$ and $L$ charges are fixed using the electric charge formula~\cite{Mohapatra:1980qe}
\begin{align}
Q \ = \ I_{3L}+I_{3R}+\frac{B-L}{2} \, .
\label{eq:charge}
\end{align}
In the scalar sector, a bi-doublet ($\phi$) and two triplets ($\Delta_{L}$, $\Delta_{R}$)
are introduced with the following quantum number assignments under ${\cal G}_{\rm LR}$: 
\begin{equation}
\phi : \: ({\bf 1},\ {\bf 2},\ {\bf 2},\ 0),\qquad  \Delta_{L}: \: ({\bf 1},\ {\bf 3},\ {\bf 1},\ 2),\qquad  \Delta_{R} : \: ({\bf 1},\ {\bf 1},\ {\bf 3},\ 2)\,.\label{eq:LRV}
\end{equation}
It is conventional to adopt the matrix representation in which $\phi$
and $\Delta_{L,\thinspace R}$ are written as 
\begin{equation}
\phi \ = \ \left(\begin{array}{cc}
\phi_{1}^{0} & \phi_{1}^{+}\\
\phi_{2}^{-} & \phi_{2}^{0}
\end{array}\right),\qquad  
\Delta_{L} \ = \ \left(\begin{array}{cc}
\delta_{L}^{+}/\sqrt{2} & \delta_{L}^{++}\\
\delta_{L}^{0} & -\delta_{L}^{+}/\sqrt{2}
\end{array}\right),\qquad 
\Delta_{R} \ = \ \left(\begin{array}{cc}
\delta_{R}^{+}/\sqrt{2} & \delta_{R}^{++}\\
\delta_{R}^{0} & -\delta_{R}^{+}/\sqrt{2}
\end{array}\right).\label{eq:LRV-2}
\end{equation}
The corresponding transformation rules are
\begin{align}
SU(2)_{L}\otimes SU(2)_{R}:\qquad & \phi \ \rightarrow \ U_{L}\phi U_{R}^{\dagger}, \quad  \Delta_{L} \ \rightarrow \  U_{L}\Delta_{L}U_{L}^{\dagger},\quad  \Delta_{R} \ \rightarrow \ U_{R}\Delta_{R}U_{R}^{\dagger}\,,\label{eq:LRV-3} \\
U(1)_{B-L}:\qquad & \phi \ \rightarrow \ \phi,  \quad \Delta_{L}\ \rightarrow \ e^{i\theta_{B-L}}\Delta_{L}, \quad   \Delta_{R} \ \rightarrow \ e^{i\theta_{B-L}}\Delta_{R}\,,\label{eq:LRV-3-1}
\end{align}
for $U_{L}\in SU(2)_{L}$, $U_{R}\in SU(2)_{R}$ and $e^{i\theta_{B-L}}\in U(1)_{B-L}$.
Note that $\tilde{\phi}\equiv\sigma_{2}\phi^{*}\sigma_{2}=-\epsilon\phi^{*}\epsilon$
transforms in the same way as $\phi$. 

The model also has a discrete left-right symmetry, which can either be 
the ${\cal P}$ parity or the ${\cal C}$ parity:
\begin{eqnarray}
{\cal P}: \qquad & \  & \phi \ \rightarrow \ \phi^{\dagger},\quad \Delta_{L} \ \leftrightarrow \ \Delta_{R}\,,\label{eq:LRV-12}\\
{\cal C}: \qquad & \  & \phi \ \rightarrow \ \phi^{T},\quad \Delta_{L} \ \leftrightarrow \ \Delta_{R}^{*}\,.
\end{eqnarray}
The scalar potential with ${\cal P}$ parity is more constrained
than that with ${\cal C}$ parity\footnote{See Eqs.~(9) and (10) in Ref.~\cite{Maiezza:2016ybz} for comparison.}
as the latter allows several complex phases. 
In this paper, for simplicity, we assume all the couplings in the scalar potential to be real, i.e.\ there is no explicit CP violation\footnote{We do, however, find that a small portion of the randomly generated samples in Sec.~\ref{sec:numerical}  have complex VEVs though the potential parameters are real, which implies that spontaneous CP violation is possible in this model. See Refs.~\cite{Branco:1985ng, Basecq:1985sx, Gunion:1989in, Deshpande:1990ip, Barenboim:2001vu, Kiers:2005gh} for more details.} in the potential. Such a potential respects both  parities. 



In general the full potential contains 17 gauge invariant terms \cite{Deshpande:1990ip}. 
 After spontaneous symmetry breaking, some components of $\phi$
and $\Delta_{L, R}$  obtain nonzero vacuum expectation values (VEVs)
while the others do not, depending on the parameters of the potential.
Since the gauge symmetry $SU(2)_{L}\otimes SU(2)_{R}\otimes U(1)_{B-L}$
is required to break to $U(1)_{{\rm em}}$ by these scalar fields,
the desired VEV alignment is \cite{Deshpande:1990ip}
\begin{equation}
\langle\phi\rangle \ = \ \frac{1}{\sqrt{2}}\left(\begin{array}{cc}
\kappa_{1} & 0\\
0 & \kappa_{2}e^{i\theta_{2}}
\end{array}\right),\qquad  \langle\Delta_{L}\rangle \ = \ \frac{1}{\sqrt{2}}\left(\begin{array}{cc}
0 & 0\\
v_{L}e^{i\theta_{L}} & 0
\end{array}\right),\qquad \langle\Delta_{R}\rangle \ = \ \frac{1}{\sqrt{2}}\left(\begin{array}{cc}
0 & 0\\
v_{R} & 0
\end{array}\right).\label{eq:LRV-1}
\end{equation}
The VEVs should furthermore obey the hierarchy $v_{L}\ll\kappa_{1,\thinspace2}\ll v_{R}$
($v_{L}$ may vanish) to meet the known phenomenology,
such as tiny neutrino masses, heavy RH gauge boson masses, the electroweak precision
parameter $\rho\simeq 1$, etc. 

Although Eq.~(\ref{eq:LRV-1}) is what we need to successfully achieve spontaneous symmetry breaking in  the LRSM, for general (arbitrary)
values of the parameters,  the scalar potential does not
necessarily lead to this VEV alignment. For
example, if we minimize the scalar potential we may obtain a minimum
with nonzero diagonal VEVs of $\langle\Delta_{L}\rangle$ or $\langle\Delta_{R}\rangle$,
which would break $U(1)_{{\rm em}}$. It is also possible to get a
minimum with $\langle\Delta_{L}\rangle=\langle\Delta_{R}\rangle$
which would imply unbroken parity symmetry.  The various
possibilities of symmetry breaking with the full scalar potential of
LRSM, due to the considerable complexity, have never been comprehensively
studied before. 

In this paper, we will therefore address an essential question of the spontaneous
symmetry breaking in  the LRSM: 

\vspace{0.0cm}
\begin{center}

\emph{How can we obtain the VEV alignment in Eq.~(\ref{eq:LRV-1}) and how likely is this?}

\end{center}
\vspace{0.0cm}

Since the full scalar potential is very complicated, a purely analytical
study is difficult and we mainly adopt a numerical approach. However, we provide 
some illustrative analytical studies for simplified cases, 
where a lot of terms in the potential are absent. Nevertheless, the analytical results give us some useful 
insight into the vacuum structure of the scalar potential and serve as a supplement to the full numerical calculations. Our numerical approach has already been established 
in Refs.~\cite{Xu:2016klg,Xu:2017vpq,Chen:2018uim} to successfully analyze beyond the SM 
scalar potentials. In general, given specific values of the potential
parameters, we can always use a computer program to numerically minimize
the potential and obtain a minimum. With further developed algorithms (see the details presented in Sec.~\ref{sec:numerical}),
we can make the program capable of identifying the zero entries in Eq.~(\ref{eq:LRV-1}).
In this way we can find out all possible
VEV alignments that can be obtained in the LRSM potential. 
We choose not to use the {\tt Vevacious} package \cite{Camargo-Molina:2013qva}, which, among other things, can also provide the minima of beyond the SM scalar potentials. This package currently has limited capability to the case we are interested in, since our potential contains many parameters and field components. In addition, we will perform a statistical analysis with a large number of random samples.  Hence, we use a self-written dedicated minimization program, which we have made publicly available in {\tt GitHub}~\cite{github}.    

As noted earlier, the significance of such a study is two-fold. First of all, for any
given set of potential parameters, we can infer whether it can lead
to successful symmetry breaking and whether the minimum of the potential is a global minimum. This in turn can put
constraints on the potential parameters and also on the scalar mass spectrum. If the LRSM is taken as a
serious theory of particle interactions beyond the SM, then these theoretical constraints from vacuum stability  should be taken into consideration, in combination with other theoretical constraints, such as unitarity and perturbativity~\cite{Maiezza:2016bzp, Chakrabortty:2016wkl, CDMZ}, as well as experimental constraints from lepton flavor violation, neutrinoless double beta decay, rare meson decays and colliders~\cite{Tello:2010am, Nemevsek:2011hz, Das:2012ii, Chen:2013fna, Dev:2013vxa, Dev:2015kca, Bambhaniya:2015ipg, Castillo-Felisola:2015bha, Lindner:2016lpp, Lindner:2016lxq, Bonilla:2016fqd, Nemevsek:2016enw, FileviezPerez:2017zwm, Dev:2017dui, Mandal:2017tab, Nemevsek:2018bbt, Barry:2012ga,Dev:2018kpa}. 

Secondly, for extended LRSMs with modified scalar sectors, e.g.\ with additional
scalar bi-doublets, triplets, singlets, etc.~\cite{Senjanovic:1978ev, Chang:1984uy, Brahmachari:2003wv, Malinsky:2005bi, Dev:2009aw, Holthausen:2009uc, Chakrabortty:2010zk, Borah:2010zq, Heeck:2015qra, Rodejohann:2015hka, Brdar:2018sbk, Majumdar:2018eqz}, one would again be concerned about the question of whether the desired VEVs can be obtained. While the potential for such cases
may be too complicated to repeat the analytical calculations given here, our
numerical method can be easily implemented to analyze the vacuum structures of such models. We leave these studies for future work.

\section{The scalar potential \label{sec:basic}}
\noindent 
The most general gauge invariant scalar potential invariant under Eq.~(\ref{eq:LRV-3})
 contains 17 independent terms: 
\begin{eqnarray}
V & \ = \  & -\mu_{1}^{2}\text{Tr}[\phi^{\dagger}\phi]-\mu_{2}^{2}\left(\text{Tr}[\tilde{\phi}\phi^{\dagger}]+\text{Tr}[\tilde{\phi}^{\dagger}\phi]\right)-\mu_{3}^{2}\left(\text{Tr}[\Delta_{L}\Delta_{L}^{\dagger}]+\text{Tr}[\Delta_{R}\Delta_{R}^{\dagger}]\right)+\lambda_{1}\text{Tr}[\phi^{\dagger}\phi]^{2}\nonumber \\
 &  & +\lambda_{2}\left(\text{Tr}[\tilde{\phi}\phi^{\dagger}]^{2}+\text{Tr}[\tilde{\phi}^{\dagger}\phi]^{2}\right)+\lambda_{3}\text{Tr}[\tilde{\phi}\phi^{\dagger}]\text{Tr}[\tilde{\phi}^{\dagger}\phi]+\lambda_{4}\text{Tr}[\phi^{\dagger}\phi]\left(\text{Tr}[\tilde{\phi}\phi^{\dagger}]+\text{Tr}[\tilde{\phi}^{\dagger}\phi]\right)\nonumber \\
 &  & +\rho_{1}\left(\text{Tr}[\Delta_{L}\Delta_{L}^{\dagger}]^{2}+\text{Tr}[\Delta_{R}\Delta_{R}^{\dagger}]^{2}\right)+\rho_{2}\left(\text{Tr}[\Delta_{L}\Delta_{L}]\text{Tr}[\Delta_{L}^{\dagger}\Delta_{L}^{\dagger}]+\text{Tr}[\Delta_{R}\Delta_{R}]\text{Tr}[\Delta_{R}^{\dagger}\Delta_{R}^{\dagger}]\right)\nonumber \\
 &  & +\rho_{3}\text{Tr}[\Delta_{L}\Delta_{L}^{\dagger}]\text{Tr}[\Delta_{R}\Delta_{R}^{\dagger}]+\rho_{4}\left(\text{Tr}[\Delta_{L}\Delta_{L}]\text{Tr}[\Delta_{R}^{\dagger}\Delta_{R}^{\dagger}]+\text{Tr}[\Delta_{L}^{\dagger}\Delta_{L}^{\dagger}]\text{Tr}[\Delta_{R}\Delta_{R}]\right) \\
 &  & +\alpha_{1}\text{Tr}[\phi^{\dagger}\phi]\left(\text{Tr}[\Delta_{L}\Delta_{L}^{\dagger}]+\text{Tr}[\Delta_{R}\Delta_{R}^{\dagger}])+\alpha_{3}(\text{Tr}[\phi\phi^{\dagger}\Delta_{L}\Delta_{L}^{\dagger}]+\text{Tr}[\phi^{\dagger}\phi\Delta_{R}\Delta_{R}^{\dagger}]\right)\nonumber \\
 &  & +\alpha_{2}\left(\text{Tr}[\Delta_{L}\Delta_{L}^{\dagger}]\text{Tr}[\tilde{\phi}\phi^{\dagger}]+\text{Tr}[\Delta_{R}\Delta_{R}^{\dagger}]\text{Tr}[\tilde{\phi}^{\dagger}\phi]+{\rm H.c.}\right) \nonumber \\
 &  & +\beta_{1}\left(\text{Tr}[\phi\Delta_{R}\phi^{\dagger}\Delta_{L}^{\dagger}]+\text{Tr}[\phi^{\dagger}\Delta_{L}\phi\Delta_{R}^{\dagger}]\right)+\beta_{2}\left(\text{Tr}[\tilde{\phi}\Delta_{R}\phi^{\dagger}\Delta_{L}^{\dagger}]+\text{Tr}[\tilde{\phi}^{\dagger}\Delta_{L}\phi\Delta_{R}^{\dagger}]\right)\nonumber \\
 &  & +\beta_{3}\left(\text{Tr}[\phi\Delta_{R}\text{\ensuremath{\tilde{\phi}^{\dagger}\Delta_{L}^{\dagger}}}]+\text{Tr}[\phi^{\dagger}\Delta_{L}\text{\ensuremath{\tilde{\phi}\Delta_{R}^{\dagger}}}]\right) \nonumber \,,\label{eq:LRV-4}
\end{eqnarray}
where, as we mentioned above, all couplings are assumed real. 
The spontaneous symmetry breaking in such a complicated potential
may result in various types of VEV alignments.  Some of them may
successfully achieve the desired symmetry breaking given by Eq.~\eqref{eq:LRV-1} and have phenomenologically
viable consequences whereas some may not. In what follows, for convenience,
we will refer to the former as \emph{good vacua} and the later as
\emph{bad vacua}.

\subsection{Good Vacua}
Since the electromagnetic gauge symmetry $U(1)_{{\rm em}}$ should
not be broken in any extension of the SM, only the electric neutral components
(namely $\phi_{1}^{0}$, $\phi_2^0$, $\delta_{L}^{0}$ and $\delta_{R}^{0}$)
can acquire nonzero VEVs.   In general, the charge conserving VEVs
of $\phi$, $\Delta_{L}$ and $\Delta_{R}$ should be 
\begin{equation}
\langle\phi\rangle \ = \ \frac{1}{\sqrt{2}}\left(\begin{array}{cc}
\kappa_{1}e^{i\theta_{1}} & 0\\
0 & \kappa_{2}e^{i\theta_{2}}
\end{array}\right),\qquad  \langle\Delta_{L}\rangle \ = \ \frac{1}{\sqrt{2}}\left(\begin{array}{cc}
0 & 0\\
v_{L}e^{i\theta_{L}} & 0
\end{array}\right),\qquad  \langle\Delta_{R}\rangle \ = \ \frac{1}{\sqrt{2}}\left(\begin{array}{cc}
0 & 0\\
v_{R}e^{i\theta_{R}} & 0
\end{array}\right).\label{eq:LRV-1-1}
\end{equation}
However, two of the phases (let us take $\theta_{1}$ and $\theta_{R}$)
can be removed by the transformation (\ref{eq:LRV-3}) with 
\begin{equation}
U_{L} \ = \ \left(\begin{array}{cc}
e^{-i(\theta_{1}-\frac{1}{2}\theta_{R})} & 0\\
0 & e^{i(\theta_{1}-\frac{1}{2}\theta_{R})}
\end{array}\right),\qquad  U_{R} \ = \ \left(\begin{array}{cc}
e^{i\theta_{R}/2} & 0\\
0 & e^{-i\theta_{R}/2}
\end{array}\right).\label{eq:LRV-5}
\end{equation}
Meanwhile, $\theta_{2}$ and $\theta_{L}$ are transformed to $\theta_{2}+\theta_{1}$
and $\theta_{L}+2\theta_{1}-\theta_{R}$, which for simplicity can
be redefined as $\theta_{2}$ and $\theta_{L}$. Therefore, without loss of generality, one can
always set 
\begin{equation}
\theta_{1} \ = \ \theta_{R} \ = \ 0 \, ,\label{eq:LRV-47}
\end{equation}
which reduces Eq.~(\ref{eq:LRV-1-1}) to Eq.~\eqref{eq:LRV-1}. 
Assuming the potential has a minimum for the VEVs given by Eq.~(\ref{eq:LRV-1}), one can replace the fields
with their VEVs and derive the minimization conditions:
\begin{equation}
\frac{\partial V}{\partial\kappa_{1}} \ = \ \frac{\partial V}{\partial\kappa_{2}} \ = \ \frac{\partial V}{\partial v_{L}} \ = \ \frac{\partial V}{\partial v_{R}} \ = \ \frac{\partial V}{\partial\theta_{2}} \ = \ \frac{\partial V}{\partial\theta_{L}} \ = \ 0 \, .\label{eq:LRV-6}
\end{equation}
From Eq.~(\ref{eq:LRV-6}) one can derive (see Appendix \ref{sec:seesaw})
the renowned seesaw relation of VEVs in LRSM \cite{Mohapatra:1980yp}:
\begin{equation}
\beta_{1}\kappa_{1}\kappa_{2}\cos\left(\theta_{2}-\theta_{L}\right)+\beta_{2}\kappa_{1}^{2}\cos\theta_{L}+\beta_{3}\kappa_{2}^{2}\cos\left(2\theta_{2}-\theta_{L}\right) \ = \ (2\rho_{1}-\rho_{3})v_{L}v_{R}.\label{eq:LRV-7}
\end{equation}
The left-hand side is roughly of the order $\beta v^{2}$ where $\beta=\beta_{1,\thinspace2,\thinspace3}$
and $v=246$ GeV, and the right-hand side is $\rho\, v_{L}v_{R}$ where
$\rho\equiv2\rho_{1}-\rho_{3}$.  From $\beta v^{2}=\rho \,v_{L}v_{R}$,
one can see that for very large $v_{R}$ (correspondingly very heavy
$W_{R}$), $v_{L}$  will
be suppressed by $1/v_{R}$, corresponding to very tiny neutrino masses, known as the seesaw relation of the
VEVs. 

Here we would like to give two comments. 
\begin{itemize}
\item  Eq.~(\ref{eq:LRV-7}) holds only if $v_{L}^{2}\neq v_{R}^{2}$.
If $v_{L}^{2}=v_{R}^{2}$, then the VEVs ($\kappa_{1}$, $\kappa_{2}$,
$v_{L}$, and $v_{R}$) may violate the relation (\ref{eq:LRV-7})
while the derivatives in Eq.~(\ref{eq:LRV-6}) remain zero. This can
be seen from the analytical calculations in Appendix \ref{sec:seesaw}
and is also verified in our numerical studies. Despite that $v_{L}^{2}=v_{R}^{2}$
is phenomenologically not allowed (since parity
is not broken in this case), it turns out that this case appears much more frequently
in the numerical scan than the case with $v_{L}^{2}\neq v_{R}^{2}$. Therefore, if in numerical
studies a minimum is obtained with the same VEV alignments as in Eq.~(\ref{eq:LRV-1-1}),
one should carefully check whether $v_{L}^{2}=v_{R}^{2}$. Only when $v_{L}^{2}\neq v_{R}^{2}$,
it is a good vacuum with the VEVs satisfiying the seesaw relation (\ref{eq:LRV-7}).
\item Eq.~(\ref{eq:LRV-6}) is based on the assumption that a minimum in
the form of Eq.~(\ref{eq:LRV-1-1}) exists. The six equations in (\ref{eq:LRV-6})
can (in general) always be solved with respect to the six variables ($\kappa_{1}$,
$\kappa_{2}$, $v_{L}$, $v_{R}$, $\theta_{2}$, $\theta_{L}$).
However, the existence of solutions of Eq.~(\ref{eq:LRV-6}) implies 
neither the existence of the minimum, nor that the first-order derivatives
with respect to the fields vanish, i.e. $\partial V/\partial\varphi =0$, 
where $\varphi$ stands for all components of $\phi$ and
$\Delta_{L,R}$.  
\end{itemize}

\subsection{Bad Vacua \label{sub:bad-vacuum}}

Next, we shall investigate other vacua that could appear but would
lead to unacceptable physical or phenomenological consequences, e.g.\ $U(1)_{{\rm em}}$
breaking. Without the requirement of charge conservation, in general,
any components of $\phi$ and $\Delta_{L,R}$ could acquire nonzero
VEVs. But before we put nonzero VEVs arbitrarily, we need to examine
the symmetries in the potential to avoid considering redundant cases.

First of all, there is the gauge symmetry $SU(2)_{L}\otimes SU(2)_{R}\otimes U(1)_{B-L}$,
which allows one to remove some degrees of freedom (DOF) in the scalar
fields by gauge fixing. By analogy to the SM case, where the Higgs doublet
in the unitarity gauge has only one DOF (the physical Higgs boson),
in the LRSM we can adopt a similar gauge to remove $3+3+1=7$ DOFs, equal
to the number of gauge bosons. More specifically, we use the transformations
$U_{L}$ and $U_{R}$  in Eq.~(\ref{eq:LRV-3}) to diagonalize $\phi$
(not $\langle\phi\rangle$) so that 
\begin{equation}
\phi \ = \ \left(\begin{array}{cc}
\phi_{1}e^{i\theta_{1}} & 0\\
0 & \phi_{2}e^{i\theta_{2}}
\end{array}\right),\qquad \Delta_{L} \ = \ \left(\begin{array}{cc}
\frac{b_{1}}{\sqrt{2}}e^{i\beta_{1}} & c_{1}e^{i\gamma_{1}}\\
a_{1}e^{i\alpha_{1}} & -\frac{b_{1}}{\sqrt{2}}e^{i\beta_{1}}
\end{array}\right),\qquad \Delta_{R}\ = \ \left(\begin{array}{cc}
\frac{b_{2}}{\sqrt{2}}e^{i\beta_{2}} & c_{2}e^{i\gamma_{2}}\\
a_{2}e^{i\alpha_{2}} & -\frac{b_{2}}{\sqrt{2}}e^{i\beta_{2}}
\end{array}\right).\label{eq:LRV-8}
\end{equation}
Some phases can be further removed by Eq.~(\ref{eq:LRV-5}) and $U(1)_{B-L}$
transformations\footnote{The explicit phase removing process to get Eq.~(\ref{eq:LRV-8}) is
as follows. First, $U(1)_{B-L}$ allows overall phase transformations
of $\Delta_{L}$ and $\Delta_{R}$, which can be used to make the
diagonal part of $\Delta_{R}$ real, i.e.\  $\beta_{2}=0$. Then one
applies Eq.~(\ref{eq:LRV-5}) to remove phases in the 1-1 and 2-1
entries of $\phi$ and $\Delta_{R}$ respectively. The phases of the
diagonal parts of $\Delta_{L}$ and $\Delta_{R}$ will not be changed
by Eq.~(\ref{eq:LRV-5}).} so that one can further set
\begin{equation}
\theta_{1} \ = \ \alpha_{2} \ = \ \beta_{2}\ = \ 0 \, .
\end{equation}
As one can check, indeed seven DOFs have been removed.

Secondly, a vacuum with $\langle\phi_{1}\rangle=\langle\phi_{2}\rangle$
and nonzero $\langle b_{1,\thinspace2}\rangle$ or $\langle c_{1,\thinspace2}\rangle$
does not necessarily break $U(1)_{{\rm em}}$ due to additional symmetries
in the vacuum. Note that if $\langle\phi_{1}\rangle=\langle\phi_{2}\rangle$,
then $\langle\phi\rangle$ is invariant under
\begin{equation}
\langle\phi\rangle \ \rightarrow \ U_{L}\langle\phi\rangle U_{R}^{\dagger},\qquad  U_{R} \ = \ U_{L}\left(\begin{array}{cc}
1 & 0\\
0 & e^{i\theta_{2}}
\end{array}\right),\label{eq:LRV-9}
\end{equation}
where $U_{L}$ can be any $SU(2)$ matrix. Accordingly, $\langle\Delta_{L}\rangle$
and $\langle\Delta_{R}\rangle$ will be transformed by the above $U_{L}$
and $U_{R}$ to other forms. For example, if $\theta_{2}=0$ and the following identical textures of 
$\langle\Delta_{L,R}\rangle$ are realized, 
\begin{equation}
\langle\Delta_{L}\rangle \ \propto \ \langle\Delta_{R}\rangle \ \propto \ \left(\begin{array}{cc}
1 & -1\\
1 & -1
\end{array}\right),\label{eq:LRV-10}
\end{equation}
then one could choose the following $U_{L,R}$ transformations: 
\begin{equation}
U_{L} \ = \ U_{R} \ = \ \frac{1}{\sqrt{2}}\left(\begin{array}{cc}
1 & -1\\
1 & 1
\end{array}\right).\label{eq:LRV-10}
\end{equation}
It is now straightforward to show that one obtains the same VEV alignment as in Eq.~(\ref{eq:LRV-1}). 
This implies that the vacuum in this example is physically equivalent
to the charge-conserving vacuum. 

A necessary and sufficient condition to infer whether $\langle\Delta_{L}\rangle$
and $\langle\Delta_{R}\rangle$ really break $U(1)_{{\rm em}}$ 
in the absence of $\langle \phi \rangle $
is that $U(1)_{{\rm em}}$ is not broken \emph{if
and only if} 
\begin{equation}
\det\langle\Delta_{L}\rangle \ = \ \det\langle\Delta_{R}\rangle \ = \ 0 \, .\label{eq:LRV-11}
\end{equation}
It is straightforward to see that if $U(1)_{{\rm em}}$ is not broken, 
then the determinants must be zero. The converse, however, needs a
short proof: first, note that in the absence of $\langle\phi\rangle$,
for any $\langle\Delta_{L}\rangle$ and $\langle\Delta_{R}\rangle$,
we can always transform them via the Schur decomposition\footnote{The Schur decomposition states that a arbitrary complex square matrix $A$ can always be decomposed into $A=UTU^{\dagger}$, where $U$ is a unitary matrix and $T$ is a lower triangular matrix (i.e., $T_{ij}=0$ for $i<j$)---see, e.g., \url{http://mathworld.wolfram.com/SchurDecomposition.html}. } to the following form: 
\begin{equation}
\langle\Delta_{L}\rangle \ \rightarrow \ U_{L}\langle\Delta_{L}\rangle U_{L}^{\dagger} \ = \ \frac{1}{\sqrt{2}}\left(\begin{array}{cc}
x_{L} & 0\\
v_{L} & -x_{L}
\end{array}\right),\qquad  \ \langle\Delta_{R}\rangle \ \rightarrow \  U_{R}\langle\Delta_{R}\rangle U_{R}^{\dagger} \ = \ \frac{1}{\sqrt{2}}\left(\begin{array}{cc}
x_{R} & 0\\
v_{R} & -x_{R}
\end{array}\right).\label{eq:schur}
\end{equation}
The determinants $\det\langle\Delta_{L}\rangle=x_{L}^{2}$ and
$\det\langle\Delta_{R}\rangle=x_{R}^{2}$ imply that if they are
zero, the diagonal elements in Eq.~(\ref{eq:schur}) must be zero, hence 
$U(1)_{{\rm em}}$ is conserved if one makes arbitrary $SU(2)_{L}\times SU(2)_{R}$ transformations. 

However, in the presence of $\langle\phi\rangle$ such transformations
may be partially or fully forbidden, depending on whether $\langle\phi_{1}\rangle=\langle\phi_{2}\rangle$
or not. In this case, it could be that $\langle\Delta_{L}\rangle$
and $\langle\Delta_{R}\rangle$ with zero determinants break $U(1)_{{\rm em}}$.
But in our numerical study presented in Sec.~\ref{sec:numerical},
among a large number of randomly generated samples, we do not find
any samples belonging to this exotic category. Therefore, based on a
high-statistics numerical study, we can draw the conclusion that in
the presence of $\langle\phi\rangle$, generally $\det\langle\Delta_{L}\rangle=\det\langle\Delta_{R}\rangle=0$
is sufficient to ensure the conservation of $U(1)_{{\rm em}}$.

Finally, some discrete symmetries may connect the vacuum in Eq.~(\ref{eq:LRV-1})
to other vacua. Consider the following two vacua: 
\begin{equation}
{\rm Vac.\ 1}:\ \ \ \langle\phi\rangle=\frac{1}{\sqrt{2}}\left(\begin{array}{cc}
\kappa_{2}e^{i\theta_{2}} & 0\\
0 & \kappa_{1}
\end{array}\right),\ \langle\Delta_{L}\rangle=\frac{1}{\sqrt{2}}\left(\begin{array}{cc}
0 & v_{L}e^{i\theta_{L}}\\
0 & 0
\end{array}\right),\ \langle\Delta_{R}\rangle=\frac{1}{\sqrt{2}}\left(\begin{array}{cc}
0 & v_{R}\\
0 & 0
\end{array}\right),\label{eq:LRV-13}
\end{equation}
\begin{equation}
{\rm Vac.\ 2}:\ \ \ \langle\phi\rangle=\frac{1}{\sqrt{2}}\left(\begin{array}{cc}
\kappa_{1} & 0\\
0 & \kappa_{2}e^{-i\theta_{2}}
\end{array}\right),\ \langle\Delta_{L}\rangle=\frac{1}{\sqrt{2}}\left(\begin{array}{cc}
0 & 0\\
v_{R} & 0
\end{array}\right),\ \langle\Delta_{R}\rangle=\frac{1}{\sqrt{2}}\left(\begin{array}{cc}
0 & 0\\
v_{L}e^{i\theta_{L}} & 0
\end{array}\right),\label{eq:LRV-14}
\end{equation}
As one can check, the above two minima can be transformed to each
other by $U_{L}=U_{R}=i\sigma_{2}$ combined with the ${\cal P}$
parity transformation: 
\begin{equation}
{\rm Vac.\ 1}\ \xrightarrow{\ \ U_{L}=U_{R}=i\sigma_{2},\ \&\ {\cal P}\ }\ {\rm Vac.\ 2}.\label{eq:LRV-58}
\end{equation} They can also be transformed to Eq.~(\ref{eq:LRV-1})
by either $U_{L}=U_{R}=i\sigma_{2}$ or ${\cal P}$. If the potential
has a minimum at Eq.~(\ref{eq:LRV-1}), then the above minima (\ref{eq:LRV-13})
and (\ref{eq:LRV-14}) also exist and have exactly the same potential
depth as Eq.~(\ref{eq:LRV-1}). The vacuum of Eq.~(\ref{eq:LRV-13})
breaks $U(1)_{{\rm em}}$ but it always coexists with the vacuum of
Eq.~(\ref{eq:LRV-1}). The vacuum of Eq.~(\ref{eq:LRV-14}) generated
by the parity transformation, would cause $\langle\Delta_{L}\rangle\gg\langle\Delta_{R}\rangle$
if $v_{L}\ll v_{R}$, though it does not break $U(1)_{{\rm em}}$.
Combining the transformations of both $U_{L}=U_{R}=i\sigma_{2}$ and
parity, one can get one more vacuum with the same potential depth. 

Therefore, the vacuum of Eq.~(\ref{eq:LRV-1}), if it exists, is always
accompanied by several wrong vacua which have the same potential
depth and are connected by discrete symmetries\footnote{This may cause the cosmological domain wall problem \cite{Zeldovich:1974uw}---for 
evading this problem, see e.g.\ \cite{Larsson:1996sp}.}, which are subgroups of the left-right gauge symmetry. On the other hand, if one finds a minimum corresponding to one of
these wrong vacua, 
then it implies the existence of the true vacuum.
In this sense, searching for minima of these wrong types is also useful.
This is particularly important for the numerical searches to be performed
later.

In summary, a vacuum which superficially seems to be bad may actually imply the coexistence of a good vacuum, or may itself be a good vacuum up to some continuous symmetry transformation. Considering these possibilities,
we would like to propose the following gauge independent criteria
for the good vacuum: 
\begin{equation}
\boxed{{\rm good\ vacuum\ criteria:\ }\begin{aligned}({\rm a})\  & \langle\phi\rangle\neq0;\\[3pt]
({\rm b})\  & \langle\Delta_{R}\rangle\neq0\ {\rm or}\ \langle\Delta_{L}\rangle\neq0;\\[3pt]
({\rm c})\  & \det\langle\Delta_{L}\rangle=\det\langle\Delta_{R}\rangle=0;\\[3pt]
({\rm d})\  & \langle\Delta_{L}\rangle\neq\langle\Delta_{R}\rangle.
\end{aligned}
}\label{eq:LRV-good}
\end{equation}
For the VEV alignment in Eq.~(\ref{eq:LRV-1}), one can straightforwardly
check that the above criteria are satisfied. Conversely, if a minimum of
the potential satisfies Eq.~(\ref{eq:LRV-good}), the VEVs must be of
the form in Eq.~(\ref{eq:LRV-1}) or can be transformed to Eq.~(\ref{eq:LRV-1})
under the previous mentioned symmetries.


\section{Analytical study of LR vacua in limiting cases\label{sec:analytical}}


\noindent In principle, we can analytically compute the first-order derivatives
of the scalar potential to find out the minima. However, for the full
potential in Eq.~(\ref{eq:LRV-4}) the calculations are too
complicated to perform analytically. In this section, we focus on
some simplified cases in which  several terms in the potential are absent. Although as such
it is not a full analysis (and sometimes  even unrealistic), the analytical
results obtained in this way provide crucial insight into the vacuum
structure of the scalar potential. The details are expanded below
and the results are summarized at the end of this section.

The first simplification we will make is to set $\alpha$ and $\beta$
to be zero because only the $\alpha$- and $\beta$-terms ``lock" the bidoublet
with the  triplets. With $\alpha=\beta=0$, the potential can be written
as two separate parts 
\begin{equation}
V|_{\alpha=\beta=0}=V_{\phi}+V_{\Delta},\label{eq:LRV-16}
\end{equation}
where $V_{\phi}$ and $V_{\Delta}$ contain only $\phi$ and $\Delta_{L/R}$ respectively.
In this case, the global symmetry of the potential becomes much larger i.e.
$G=[SU(2)_L\times SU(2)_R]_\phi\times[SU(2)_L\times SU(2)_R]_\Delta\times U(1)_{B-L}$. (In fact this is true as long as $\alpha_3=0$ and $\beta_i=0$)
Once the $\langle\phi\rangle$ and $\langle\Delta\rangle$ vevs are switched on, the resulting symmetry will be $U(1)_L\times U(1)_R\times U(1)_Y$.
This will lead to 10 massless states out of which six will be absorbed as longitudinal modes of the gauge bosons of the theory leaving four massless states.
Clearly therefore, this theory is not realistic. But nevertheless we study the vacuum structure and the symmetries of the vacuum
in order to understand the same for the full model with all couplings turned on.

Expressed in terms of the explicit
components defined in Eq.~(\ref{eq:LRV-8}), $V_{\phi}$ and $V_{\Delta}$ are
\begin{eqnarray}
V_{\phi} & = & -\mu_{1}^{2}(\phi_{1}^{2}+\phi_{2}^{2})-4\mu_{2}^{2}\phi_{1}\phi_{2}\cos\theta_{2}\nonumber \\
 &  & +\lambda_{1}(\phi_{1}^{2}+\phi_{2}^{2})^{2}+\lambda_{2}8\phi_{1}^{2}\phi_{2}^{2}\cos2\theta_{2}+\lambda_{3}4\phi_{1}^{2}\phi_{2}^{2}+\lambda_{4}4\phi_{1}\phi_{2}\left(\phi_{1}^{2}+\phi_{2}^{2}\right)\cos\theta_{2}\,,\label{eq:LRV-17}
\end{eqnarray}
\begin{eqnarray}
V_{\Delta} & = & -\mu_{3}^{2}\left[\left(a_{1}^{2}+b_{1}^{2}+c_{1}^{2}\right)+\left(a_{2}^{2}+b_{2}^{2}+c_{2}^{2}\right)\right]\nonumber \\
 &  & +\rho_{1}\left[\left(a_{1}^{2}+b_{1}^{2}+c_{1}^{2}\right){}^{2}+\left(a_{2}^{2}+b_{2}^{2}+c_{2}^{2}\right){}^{2}\right]\nonumber \\
 &  & +\rho_{3}\left(a_{1}^{2}+b_{1}^{2}+c_{1}^{2}\right)\left(a_{2}^{2}+b_{2}^{2}+c_{2}^{2}\right) \\
 &  & +\rho_{2}\left[4a_{1}^{2}c_{1}^{2}+b_{1}^{4}+4a_{1}b_{1}^{2}c_{1}\cos\left(\alpha_{1}-2\beta_{1}+\gamma_{1}\right)+(1\rightarrow2)\right]\nonumber \\
 &  & +\rho_{4}\left[4a_{2}b_{1}^{2}c_{2}\cos\left(\alpha_{2}-2\beta_{1}+\gamma_{2}\right)+4a_{1}b_{2}^{2}c_{1}\cos\left(\alpha_{1}-2\beta_{2}+\gamma_{1}\right)\right.\nonumber \\
 &  & \ \ \ \ \ \left.+8a_{1}a_{2}c_{1}c_{2}\cos\left(\alpha_{1}-\alpha_{2}+\gamma_{1}-\gamma_{2}\right)+2b_{2}^{2}b_{1}^{2}\cos2\left(\beta_{1}-\beta_{2}\right)\right].\label{eq:LRV-18}\nonumber
\end{eqnarray}
Here the field components (e.g., $\phi_{1,\,2}$,  $a_{1,\,2}$, $b_{1,\,2}$, $c_{1,\,2}$, $\cdots$) have been defined in Eq.~(\ref{eq:LRV-8}).
To proceed, let us further set $\mu_{2}^{2}$, $\lambda_{2}$,
$\lambda_{4}$, $\rho_{2}$, and $\rho_{4}$ to zero so that we are
not bothered by the cosines appearing in the above expression.
In this very simplified
case, we note that $V_{\phi}$ and $V_{\Delta}$ have essentially
the same form:
\begin{equation}
V_{\phi0}=-\mu_{1}^{2}(\phi_{1}^{2}+\phi_{2}^{2})+\lambda_{1}(\phi_{1}^{2}+\phi_{2}^{2})^{2}+4\lambda_{3}\phi_{1}^{2}\phi_{2}^{2}\,,\label{eq:LRV-19}
\end{equation}
\begin{equation}
V_{\Delta0}=-\mu_{3}^{2}\left(\delta_{1}^{2}+\delta_{2}^{2}\right)+\rho_{1}\left(\delta_{1}^{2}+\delta_{2}^{2}\right){}^{2}+(\rho_{3}-2\rho_{1})\delta_{1}^{2}\delta_{2}^{2}\,,\label{eq:LRV-20}
\end{equation}
with 
$\delta_{1}^{2}\equiv a_{1}^{2}+b_{1}^{2}+c_{1}^{2}$ and $\delta_{2}^{2}\equiv a_{2}^{2}+b_{2}^{2}+c_{2}^{2}$. 
Here we have extracted $2\rho_{1}\delta_{1}^{2}\delta_{2}^{2}$ in
Eq.~(\ref{eq:LRV-20}) so that $V_{\Delta0}$ shows explicitly  the
same form as $V_{\phi0}$. In the following discussion, we will focus
on $V_{\phi0}$ while the conclusions can be easily transferred to
$V_{\Delta0}$ by replacing $\mu_{1}^{2}\rightarrow\mu_{3}^{2}$,
$\lambda_{1}\rightarrow\rho_{1}$ and $4\lambda_{3}\rightarrow(\rho_{3}-2\rho_{1})$.

\begin{figure}
\centering

\includegraphics[width=12cm]{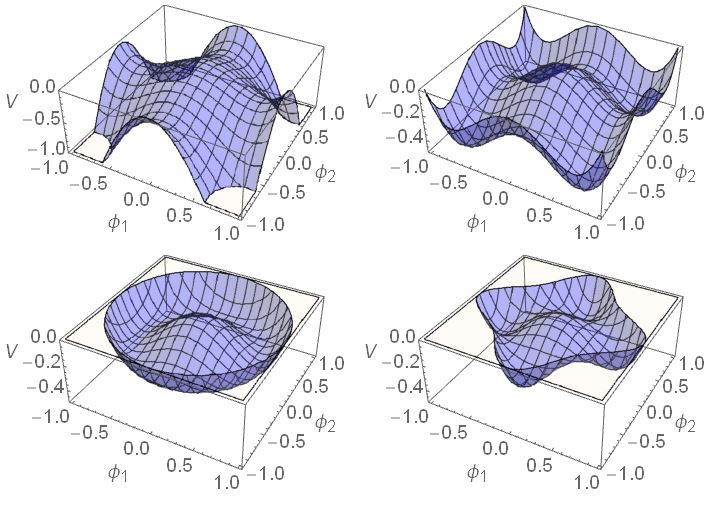}\caption{\label{fig:V-z4}The $D_{4}$-symmetric potential in Eq.~(\ref{eq:LRV-19})
for different values of $\lambda_{3}/\lambda_{1}$. In the upper left
panel ($\lambda_{3}/\lambda_{1}=-1$), the potential is not bounded
from below. In the upper right panel ($\lambda_{3}/\lambda_{1}=-1/2$),
the potential has four minima at $(\phi_{1},\ \phi_{2})=\frac{1}{\sqrt{2}}(\pm1,\ \pm1)$.
In the lower left panel ($\lambda_{3}/\lambda_{1}=0$), the potential
has infinite minima connected by an $SO(2)$ symmetry. In the lower
right panel ($\lambda_{3}/\lambda_{1}=1$), the potential has four
minima at $(\phi_{1},\ \phi_{2})=\frac{1}{\sqrt{2}}(\pm1,\ \pm1)$.}
\end{figure}

A notable feature of $V_{\phi0}$ is that it respects the following
dihedral ($D_{4}$) symmetry
\begin{equation}
D_{4}:\ (\phi_{1},\thinspace\phi_{2})^{T}\rightarrow R(\phi_{1},\thinspace\phi_{2})^{T}\,,\label{eq:LRV-23}
\end{equation}
\begin{equation}
R=\left(\begin{array}{cc}
0 & \pm1\\
\pm1 & 0
\end{array}\right),\ {\rm or}\ \left(\begin{array}{cc}
\pm1 & 0\\
0 & \pm1
\end{array}\right),\label{eq:LRV-24}
\end{equation}
which leads to $D_{4}$-symmetric vacuum structures shown in Fig.~\ref{fig:V-z4}.  
The bounded from below (BFB) condition for the potential in Eq.~(\ref{eq:LRV-19}) is manifest:
\begin{equation}
\lambda_{1}>0,\ \lambda_{3}>-\lambda_{1}.\label{eq:LRV-25}
\end{equation}
In the following discussion of minima, by default we assume the BFB
condition should be satisfied. 

From the first-order derivatives 
\begin{eqnarray}
\frac{\partial V_{\phi0}}{\partial\phi_{1}} & = & 2\phi_{1}\left[-\mu_{1}^{2}+2\lambda_{1}(\phi_{1}^{2}+\phi_{2}^{2})+4\lambda_{3}\phi_{2}^{2}\right]=0\,,\label{eq:LRV-26}\\
\frac{\partial V_{\phi0}}{\partial\phi_{2}} & = & 2\phi_{2}\left[-\mu_{1}^{2}+2\lambda_{1}(\phi_{1}^{2}+\phi_{2}^{2})+4\lambda_{3}\phi_{1}^{2}\right]=0\,,\label{eq:LRV-27}
\end{eqnarray}
we get three possible solutions
\begin{equation}
(\phi_{1}^{2},\ \phi_{2}^{2})=\frac{\mu_{1}^{2}}{4(\lambda_{1}+\lambda_{3})}(1,\ 1),\ {\rm with}\ V_{\phi0}=-\frac{\mu_{1}^{4}}{4(\lambda_{1}+\lambda_{3})},\label{eq:LRV-28}
\end{equation}
\begin{equation}
(\phi_{1}^{2},\ \phi_{2}^{2})=\frac{\mu_{1}^{2}}{2\lambda_{1}}(0,\ 1)\ {\rm or}\ \frac{\mu_{1}^{2}}{2\lambda_{1}}(1,\ 0),\ {\rm with}\ V_{\phi0}=-\frac{\mu_{1}^{4}}{4\lambda_{1}},\label{eq:LRV-29}
\end{equation}
\begin{equation}
(\phi_{1}^{2},\ \phi_{2}^{2})=(0,\ 0),\ {\rm with}\ V_{\phi0}=0.\label{eq:LRV-30}
\end{equation}
Here all the denominators such as $4(\lambda_{1}+\lambda_{3})$ and
$2\lambda_{1}$ are positive according to the BFB condition (\ref{eq:LRV-25}),
which implies that the solutions (\ref{eq:LRV-28}) and (\ref{eq:LRV-29}) will not exist if $\mu_{1}^{2}<0$,
because this will lead to negative $\phi_{1}^{2}$ or $\phi_{1}^{2}$.
Therefore, for $\mu_{1}^{2}<0$, the potential has only one minimum,
which is necessarily the global minimum of the potential.

By comparing the potential values $V_{\phi0}$ of the three solutions,
we can infer which one can be the global minimum without computing
the second-order derivatives (Hessian matrix). One can see from Eq.~(\ref{eq:LRV-28})
and (\ref{eq:LRV-29}) that if $\lambda_{3}>0$ (assuming $\mu_{1}^{2}>0$
and $\lambda_{1}>0$), then Eq.~(\ref{eq:LRV-29}) should be the global
minimum because it is deeper than the other candidates. If $\lambda_{3}<0$,
Eq.~(\ref{eq:LRV-28}) is the global  minimum. 

Actually, as one can check that if $\lambda_{3}<0$ the Hessian matrix
for Eq.~(\ref{eq:LRV-28}) is positive definite while the Hessian
matrix for Eq.~(\ref{eq:LRV-29}) loses positive definiteness, and
vice versa. In the critical case $\lambda_{3}=0$, there is an $SO(2)$
symmetry so any point on the circle $\phi_{1}^{2}+\phi_{2}^{2}=\frac{\mu_{1}^{2}}{2\lambda_{1}}$
is a global minimum.

In summary, assuming $\mu_{1}^{2}>0$  and $\lambda_{1}>0$, the
vacuum structure depends on the ratio $\lambda_{3}/\lambda_{1}$ as
follows:
\begin{itemize}
\item $\lambda_{3}/\lambda_{1}\leq-1$: $V_{\phi0}$ is not BFB;
\item $-1<\lambda_{3}/\lambda_{1}<0$: $V_{\phi0}$ has four minima at $(\phi_{1},\ \phi_{2})\propto(\pm1,\ \pm1)$,
with equal depth;
\item $\lambda_{3}/\lambda_{1}=0$: $V_{\phi0}$ has infinite minima on
the circle $\phi_{1}^{2}+\phi_{2}^{2}=\frac{\mu_{1}^{2}}{2\lambda_{1}}$,
with equal depth;
\item $\lambda_{3}/\lambda_{1}>0$: $V_{\phi0}$ has four minima at $(\phi_{1},\ \phi_{2})\propto(0,\ \pm1)$
and $(\pm1,\ 0)$, with equal depth.
\end{itemize}
In Fig.~\ref{fig:V-z4}, we show how the potential varies for different
values of $\lambda_{3}/\lambda_{1}$. We set $\mu_{1}^{2}=1$ and
$\lambda_{1}=1$ in Eq.~(\ref{eq:LRV-19}) and select four values
$-1$, $-1/2$, 0, and $1$ for $\lambda_{3}/\lambda_{1}$ so that
all the four cases above are covered.

\begin{figure}
\centering

\includegraphics[width=0.96\textwidth]{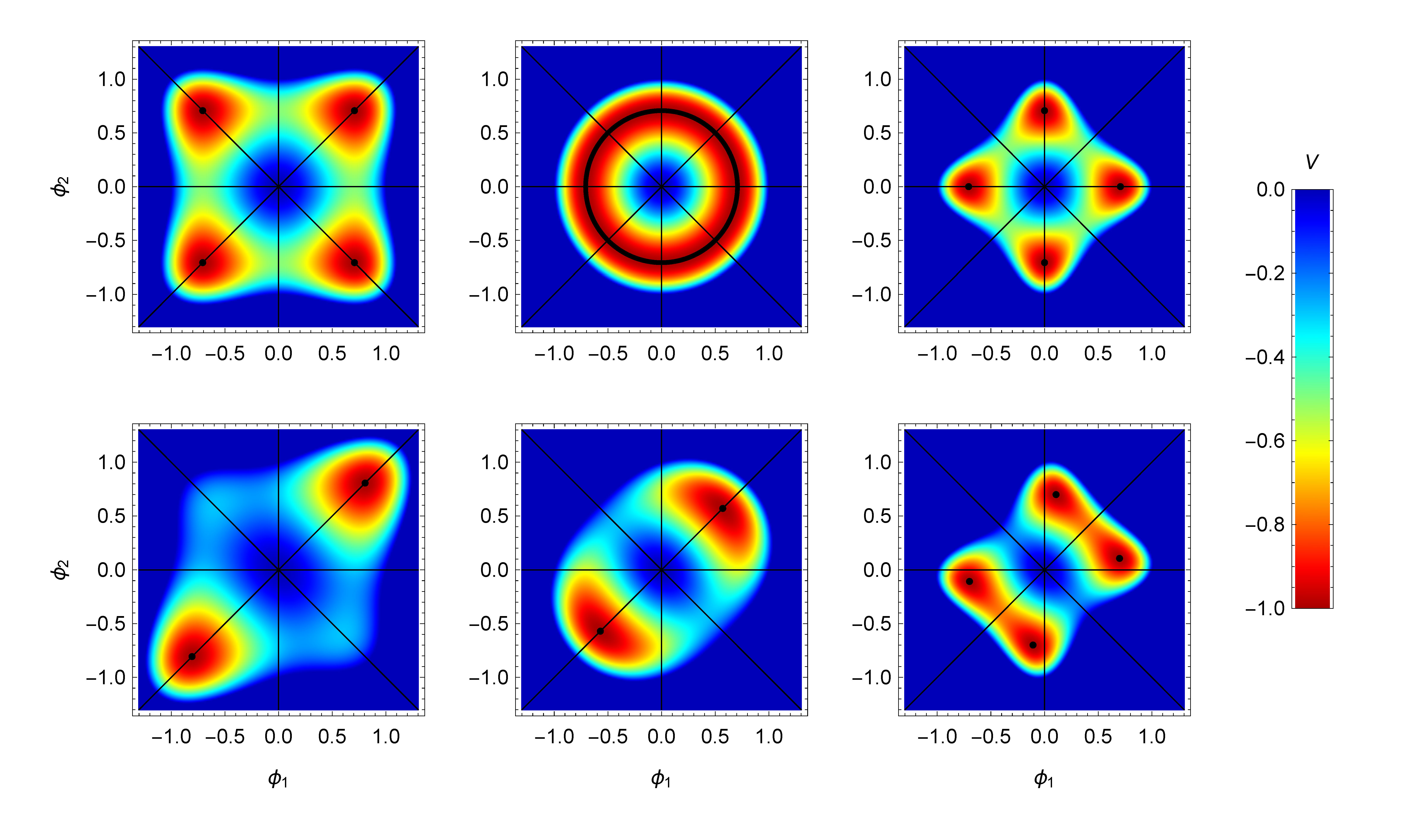}\caption{\label{fig:Vdensity} Vacua in the $D_{4}$-symmetric (upper panels)
and $D_{4}$-broken (lower panels) potentials. The $D_{4}$-symmetric
potential is given by Eq.~(\ref{eq:LRV-19}) with $\lambda_{3}=-1/2$,
0, and $1$ (from left to right) and $\mu_{1}^{2}=\lambda_{1}=1$.
The $D_{4}$-broken broken potential is obtained by adding $-4\mu_{2}^{2}\phi_{1}\phi_{2}$
to Eq.~(\ref{eq:LRV-19}) with $\mu_{2}^{2}=0.15$. This changes the
locations of stable vacua (marked in  black). The potentials
are normalized so that $V_{{\rm min}}=-1$.}

\end{figure}

So far, we have not taken $\mu_{2}^{2}$, $\lambda_{2}$, $\lambda_{4}$
into consideration. In the presence of these terms, the $D_{4}$ symmetry
will be broken and the complexity of  the above analysis will be 
increased. We do not plan to derive the corresponding analytic
expressions. But we can discuss qualitatively on the consequences
using the $D_{4}$-symmetric conclusions. If we set $\mu_{2}^{2}>0$
while keeping $\lambda_{2}$ and $\lambda_{4}$ still at zero, this is
the same as adding the term $-4\mu_{2}^{2}\phi_{1}\phi_{2}\cos\theta_{2}$
to the $D_{4}$-symmetric potential $V_{\phi0}$. From the potential
$V_{\phi0}-4\mu_{2}^{2}\phi_{1}\phi_{2}\cos\theta_{2}$ we can immediately
see that $\cos\theta_{2}$ should be $1$ (if $4\mu_{2}^{2}\phi_{1}\phi_{2}>0$)
or $-1$ (if $4\mu_{2}^{2}\phi_{1}\phi_{2}<0$) to reach a minimum,
which implies that we would still get real solutions even if we turn
on the complex phase. Therefore instead of $V_{\phi0}-4\mu_{2}^{2}\phi_{1}\phi_{2}\cos\theta_{2}$
we can focus on $V_{\phi0}-4\mu_{2}^{2}\phi_{1}\phi_{2}$. In Fig.~\ref{fig:Vdensity},
we plot both $V_{\phi0}$ and $V_{\phi0}-4\mu_{2}^{2}\phi_{1}\phi_{2}$
to show the changes caused by the $\mu_{2}^{2}$ term. In the upper
left panel ($\lambda_{3}/\lambda_{1}=-1/2$, $\mu_{2}^{2}=0$), the
four minima has the same depth due to the $D_{4}$ symmetry. When
the $\mu_{2}^{2}$ term is added (for illustration we choose $\mu_{2}^{2}=0.15$), as shown in
the corresponding lower panel, two of the minima become deeper than
the other two. Note that in this case $(\phi_{1},\thinspace\phi_{2})$
at the minima still align in the direction $(1,\thinspace1)$ or $(1,\thinspace-1)$.
In the middle panels ($\lambda_{3}/\lambda_{1}=0$), the $SO(2)$
symmetry is broken when $\mu_{2}$ is nonzero, leading also to the
VEV alignment $(\phi_{1},\thinspace\phi_{2})\propto(1,\thinspace1)$.
In the right panels ($\lambda_{3}/\lambda_{1}=1$), the four minima
still have equal depth after adding the $\mu_{2}^{2}$ term, but the
VEV alignment is changed from $(1,\thinspace0)$ or $(0,\thinspace1)$
to $(1,\thinspace r)$ or $(r,\thinspace1)$ where $r$ depends on
$\mu_{2}$.

Therefore, within the simple $D_{4}$-soft-broken potential $V_{\phi0}-4\mu_{2}^{2}\phi_{1}\phi_{2}\cos\theta_{2}$,
we can already get an arbitrary VEV alignment $(\phi_{1},\thinspace\phi_{2})\propto(1,\thinspace r)$.
Further turning on $\lambda_{2}$ and $\lambda_{4}$ couplings may
produce more possibilities (e.g.\ spontaneous CP breaking) which should
include the VEV alignments obtained in $V_{\phi0}-4\mu_{2}^{2}\phi_{1}\phi_{2}\cos\theta_{2}$.\\

Now let us discuss on the triplet sector. 
As previously mentioned, if $\rho_{2}=\rho_{4}=0$, the potential
of $\Delta_{L}$ and $\Delta_{R}$ reduces to  $V_{\Delta0}$ which
has the same form as $V_{\phi0}$. Using the previous conclusions
on $V_{\phi0}$, we know that when 
\begin{equation}
\mu_{3}^{2}>0,\ \rho_{3}>2\rho_{1}>0,\label{eq:LRV-31}
\end{equation}
the $D_{4}$-symmetric potential $V_{\Delta0}$ has four minima at
$(\delta_{1},\ \delta_{2})\propto(0,\ 1)$ or $(1,\ 0)$, which implies
the VEV alignment 
\begin{equation}
\langle\Delta_{L}\rangle=0,\ \langle\Delta_{R}\rangle\neq0\label{eq:LRV-32}
\end{equation}
can be obtained. 

We would like to point it out that even if $\rho_{2}$ and $\rho_{4}$
are nonzero, $V_{\Delta}$ is still $D_{4}$-symmetric. The symmetry
transformation is similar to Eq.~(\ref{eq:LRV-23}) with $(\phi_{1},\ \phi_{2})$
replaced by $(\Delta_{L},\ \Delta_{R})$. Consequently, the VEV alignments
of $(\Delta_{L},\ \Delta_{R})$ from Eq.~(\ref{eq:LRV-18}) have only
three possibilities: (i) one of $(\Delta_{L},\ \Delta_{R})$ is nonzero
and the other is zero, like Eq.~(\ref{eq:LRV-32}); (ii) both are
zero; or (iii) both are nonzero but $\Delta_{L}=\Delta_{R}$. 

However, the above arguments treat $\Delta_{L}$ and $\Delta_{R}$
as two singlets and do not take the fact that they have internal components
into consideration.  Analyses in terms of the components $(a_{i},\ b_{i},\ c_{i})$
will be much more complicated. We adopt a numerical method to check
the above conclusions and find that there are indeed only the three
cases, except that (iii) should be interpreted as $a_{1}^{2}+b_{1}^{2}+c_{1}^{2}=a_{2}^{2}+b_{2}^{2}+c_{2}^{2}$.

Besides, one should notice that for nonzero $\rho_{2}$ and $\rho_{4}$,
Eq.~(\ref{eq:LRV-31}) is no longer the condition to get Eq.~(\ref{eq:LRV-32}).
But at least the simple conclusion with $\rho_{2}=\rho_{4}=0$ is
enough to show that the VEV alignment in Eq.~(\ref{eq:LRV-32}) can
be obtained in a part of the parameter space.

Eq.~(\ref{eq:LRV-32}) is what we need to achieve spontaneous parity
breaking. The other two possible cases, namely $\Delta_{L}=\Delta_{R}\neq0$
and $\Delta_{L}=\Delta_{R}=0$, can be modified by $D_{4}$-breaking
terms {[}e.g.\ the $\alpha$ and $\beta$ terms in~(\ref{eq:LRV-4}){]}
so that $\Delta_{L}\neq\Delta_{R}$ and $\Delta_{R}\neq0$. However
the experimental constraints require that $\langle\Delta_{L}\rangle$
should be much smaller than $\langle\Delta_{R}\rangle$ since the
former, limited by the electroweak $\rho$ parameter, has to be lower
than the GeV scale while the latter should be above the TeV scale
(collider bounds on $W_{R}$). Such a strong hierarchy would require
substantial  fine-tuning in the scalar potential. Therefore, the
solution $\langle\Delta_{L}\rangle=0$ as a consequence of symmetry
is a more favored option.

Focused on the solution with zero $\Delta_{L}$ and nonzero $\Delta_{R}$,
we proceed to study the VEV alignments of the internal components
of $\Delta_{R}$. For simplicity, let us first set $\rho_{4}=0$ ($\mu_{3}^{2}$,
$\rho_{1}$, $\rho_{2}$, and $\rho_{3}$ are nonzero). Note that
in the remaining terms of $V_{\Delta}$, only the $\rho_{3}$ term
couples $\Delta_{L}$ to $\Delta_{R}$. This term has no contribution
to the first-order derivatives at the minimum with zero $\Delta_{L}$
because 
\[
\frac{\partial(\delta_{1}^{2}\delta_{2}^{2})}{\partial a_{i}}=\frac{\partial(\delta_{1}^{2}\delta_{2}^{2})}{\partial b_{i}}=\frac{\partial(\delta_{1}^{2}\delta_{2}^{2})}{\partial c_{i}}=0,\ \ \ {\rm for}\ \Delta_{L}=0.
\]
In the absence of $\rho_{3}$ and $\rho_{4}$, $\Delta_{R}$ decouples
with $\Delta_{L}$ in $V_{\Delta}$. Hence we only need to consider
the following part of the potential
\begin{eqnarray}
V_{\Delta2} & = & -\mu_{3}^{2}\left(a_{2}^{2}+b_{2}^{2}+c_{2}^{2}\right)+\rho_{1}\left(a_{2}^{2}+b_{2}^{2}+c_{2}^{2}\right){}^{2}+\rho_{2}\left[4a_{2}^{2}c_{2}^{2}+b_{2}^{4}+4a_{2}b_{2}^{2}c_{2}\cos\left(\alpha_{2}-2\beta_{2}+\gamma_{2}\right)\right]\label{eq:LRV-34} \nonumber \\
 & = & -\mu_{3}^{2}\delta_{2}^{2}+\rho_{1}\delta_{2}^{4}+4\rho_{2}|\det\Delta_{R}|^{2}\, ,\label{eq:LRV-36}
\end{eqnarray}
where in the second line we have simplified the $\rho_{2}$ term which
can be verified by an explicit computation. To proceed, we need the
following useful relation:  
\begin{equation}
0\leq|\det\Delta_{R}|\leq\frac{1}{2}\delta_{2}^{2}\,,\label{eq:LRV-37}
\end{equation}
which can be proven by $\delta_{2}^{2}=a_{2}^{2}+b_{2}^{2}+c_{2}^{2}\geq\max(b_{2}^{2}+2a_{2}c_{2},\ b_{2}^{2}-2a_{2}c_{2})$$\geq|b_{2}^{2}+2a_{2}c_{2}e^{i\omega}|=2|\det\Delta_{R}|$,
where $\omega=\alpha_{2}-2\beta_{2}+\gamma_{2}$. Note that for a
fixed value of $\delta_{2}^{2}$, $|\det\Delta_{R}|$ can reach any
value in the above range. Therefore, we can parametrize $|\det\Delta_{R}|$
as $\frac{1}{2}\delta_{2}^{2}\cos\theta$ and write 
\begin{equation}
V_{\Delta2}=-\mu_{3}^{2}\delta_{2}^{2}+\left(\rho_{1}+\rho_{2}\cos^{2}\theta\right)\delta_{2}^{4}\,.\label{eq:LRV-38}
\end{equation}
From Eq.~(\ref{eq:LRV-38}), it is obvious to identify the minima. Let
us take $\rho_{1}>0$ and $\mu_{3}^{2}>0$, which is necessary to
satisfy the BFB condition and obtain nonzero VEVs. If $\rho_{2}>0$,
the minimum should be at $\cos^{2}\theta=0$; and if $\rho_{2}<0$,
it should be at $\cos^{2}\theta=0$. Therefore the minima of $V_{\Delta2}$
should locate at:
\begin{equation}
\rho_{2}>0:\ \ \delta_{2}^{2}=\frac{\mu_{3}^{2}}{2\rho_{1}}\,,\ \det\Delta_{R}=0,\label{eq:LRV-41}
\end{equation}
\begin{equation}
\rho_{2}<0:\ \ \delta_{2}^{2}=\frac{\mu_{3}^{2}}{2(\rho_{1}+\rho_{2})},\ \det\Delta_{R}=\frac{1}{2}\delta_{2}^{2}=\frac{\mu_{3}^{2}}{4(\rho_{1}+\rho_{2})}\,.\label{eq:LRV-42}
\end{equation}
As long as $\alpha_{1,\thinspace2,\thinspace3}=\beta_{1,\thinspace2,\thinspace3}=0$,
we always have the freedom to transform $\Delta_{R}\rightarrow U_{R}\Delta_{R}U_{R}^{\dagger}$
individually (without the corresponding transformation of the bidoublet
$\phi$) within the triplet potential $V_{\Delta}$. According again to
the Schur decomposition, we can always transform $\Delta_{R}$
to a lower triangular matrix. In this form, Eqs.~(\ref{eq:LRV-41})
and (\ref{eq:LRV-42}) should be 
\begin{equation}
\rho_{2}>0:\ \ \Delta_{R}=\sqrt{\frac{\mu_{3}^{2}}{2\rho_{1}}}\left(\begin{array}{cc}
0 & 0\\
1 & 0
\end{array}\right),\label{eq:LRV-39}
\end{equation}
\begin{equation}
\rho_{2}<0:\ \ \Delta_{R}=\sqrt{\frac{\mu_{3}^{2}}{4(\rho_{1}+\rho_{2})}}\left(\begin{array}{cc}
1 & 0\\
0 & -1
\end{array}\right).\label{eq:LRV-40}
\end{equation}
Eq.~(\ref{eq:LRV-39}) is straightforward to get, because if $\det\Delta_{R}=0$,
only the 2-1 element can be nonzero. Eq.~(\ref{eq:LRV-40}) has zero
$a_{2}$ because when $|\det\Delta_{R}|=\frac{1}{2}\delta_{2}^{2}$,
$|a_{2}|$ and $|c_{2}|$ should be equal, according to the derivation
of Eq.~(\ref{eq:LRV-37}). 
One should keep in mind that Eqs.~(\ref{eq:LRV-39}) and (\ref{eq:LRV-40})
are derived under the assumption that $\mu_{3}^{2}>0$, $\rho_{4}=\alpha_{1,\thinspace2,\thinspace3}=\beta_{1,\thinspace2,\thinspace3}=0$, 
 and that the potential is BFB.\\

The above analyses implies the following {\it sufficient but not necessary} conditions to get a good vacuum: 
\begin{equation}
\mu_{1}^{2},\ \mu_{2}^{2},\ \mu_{3}^{2}>0,\label{eq:LRV-43}
\end{equation}
\begin{equation}
\lambda_{1}>0,\ \lambda_{2}=0,\ \lambda_{3}>-\lambda_{1},\ \lambda_{4}=0,\label{eq:LRV-44}
\end{equation}
\begin{equation}
\rho_{1}>0,\ \rho_{2}>0,\ \rho_{3}>2\rho_{1},\ \rho_{4}=0,\label{eq:LRV-45}
\end{equation}
\begin{equation}
\alpha_{1,\thinspace2,\thinspace3}=\beta_{1,\thinspace2,\thinspace3}=0.\label{eq:LRV-46}
\end{equation}
If the potential parameters satisfy the above conditions, then it can be guaranteed that the potential has a global minimum corresponding to a good vacuum. In practical use, however,  $\alpha_{i}$ and $\beta_{i}$ cannot all be set to zero because of additional massless states, as mentioned before. To solve this problem, we can add small perturbations to  $\alpha_{i}$ to avoid the massless states. 
More explicitly, using the above condition, we can easily find a set of potential parameters satisfying Eqs.~(\ref{eq:LRV-43}),  (\ref{eq:LRV-44}) and (\ref{eq:LRV-45}). Then if we set  $\alpha_{i}$ to zero, the potential has a good vacuum, though not realistic. Next we can explore the nearby around this point by tentatively adding some small perturbations to $\alpha_{i}$. If the perturbations are small enough, the conclusion should hold as well. Sometimes, the perturbations can be very large without changing the conclusion. The exploration starting   from  Eq.~(\ref{eq:LRV-46}) needs numerical assistance, as will be done in Sec.~\ref{sec:numerical}. And we will show (see Fig.~\ref{fig:a123b123}), indeed one can find some deviations  from  Eq.~(\ref{eq:LRV-46}) that lead to successful symmetry breaking.             

Besides,  we have other comments on the above conditions: 
\begin{itemize}
\item The above conditions also guarantee BFB;
\item If $-\lambda_{1}<\lambda_{3}<0$, $\langle\phi\rangle\propto{\rm diag}(1,\ 1)$;
if $0<\lambda_{3}$, $\langle\phi\rangle\propto{\rm diag}(1,\ r)$
with $r\neq1$;
\item The vacuum obtained in this way always has $\langle\Delta_{L}\rangle=0$.
\end{itemize}

The conditions (\ref{eq:LRV-43})-(\ref{eq:LRV-46}) obtained by the above analytic study will be very useful in the subsequent numerical study. It helps to quickly find out a viable region in which the potential has a good vacuum. In addition, it is also important for setting some benchmarks when studying the global minimum constraints.

\section{Numerical Study\label{sec:numerical}}
\noindent 
With the modern technology of numerical computation, given a set of
specific values of the potential parameters one can readily obtain
a numerical minimum. There have been various algorithms well developed
to find minima of multi-variable functions\footnote{See, e.g.\ \cite{nocedal2006numerical} or the {\tt SciPy} document:
\url{https://docs.scipy.org/doc/scipy/reference/generated/scipy.optimize.minimize.html#scipy.optimize.minimize}.
In this paper, we adopt the Nelder-Mead simplex method which is the
most commonly used algorithm since it does not require derivatives
of the function.}. Most algorithms are based on iterative searches which means the
program starts by given an initial point and iteratively computes
the next step according to some principles until the steps converge
to a minimum. The convergence can not be guaranteed, so for a single
process searching for minima there is a small probability of failure.
In case of failure, one can try to start the process again with a
different initial point. Repeating the processes will eventually arrive
at a minimum which may be local or global. 

Once we get a numerical minimum of the potential, we can inspect the
field component values at the minimum, checking if some of them are
zero or some of them are equal. However, in numerical calculations
it is impossible to have infinite accuracy so the would-be zero numbers
are generally nonzero but very small (e.g.\ $10^{-6}\sim10^{-9}$ for
numerical minimization based on a 8-byte real number system). This
implies that, due to limited accuracy, there is no absolute equality
in the numerical results. The simplest solution to this problem is
by setting a cut on the difference of two numbers, below which the
two numbers are thought to be equal and above which they are not\footnote{Actually, we use this to check whether a quantity is zero while for checking whether two quantities $a$ and $b$ are equal, we convert it to the problem of checking whether $(a-b)/b$ is zero.}.
But this may cause misjudgment since the difference of two actually
unequal numbers may occasionally be smaller than the cut. To double
check if a field component is zero at the  minimum, we can
invoke the minimization process again with  this component fixed
to zero. If it gives a slightly better minimization, then we can conclude
that it is zero at the minimum.

\begin{figure}
\centering

\includegraphics[width=10cm]{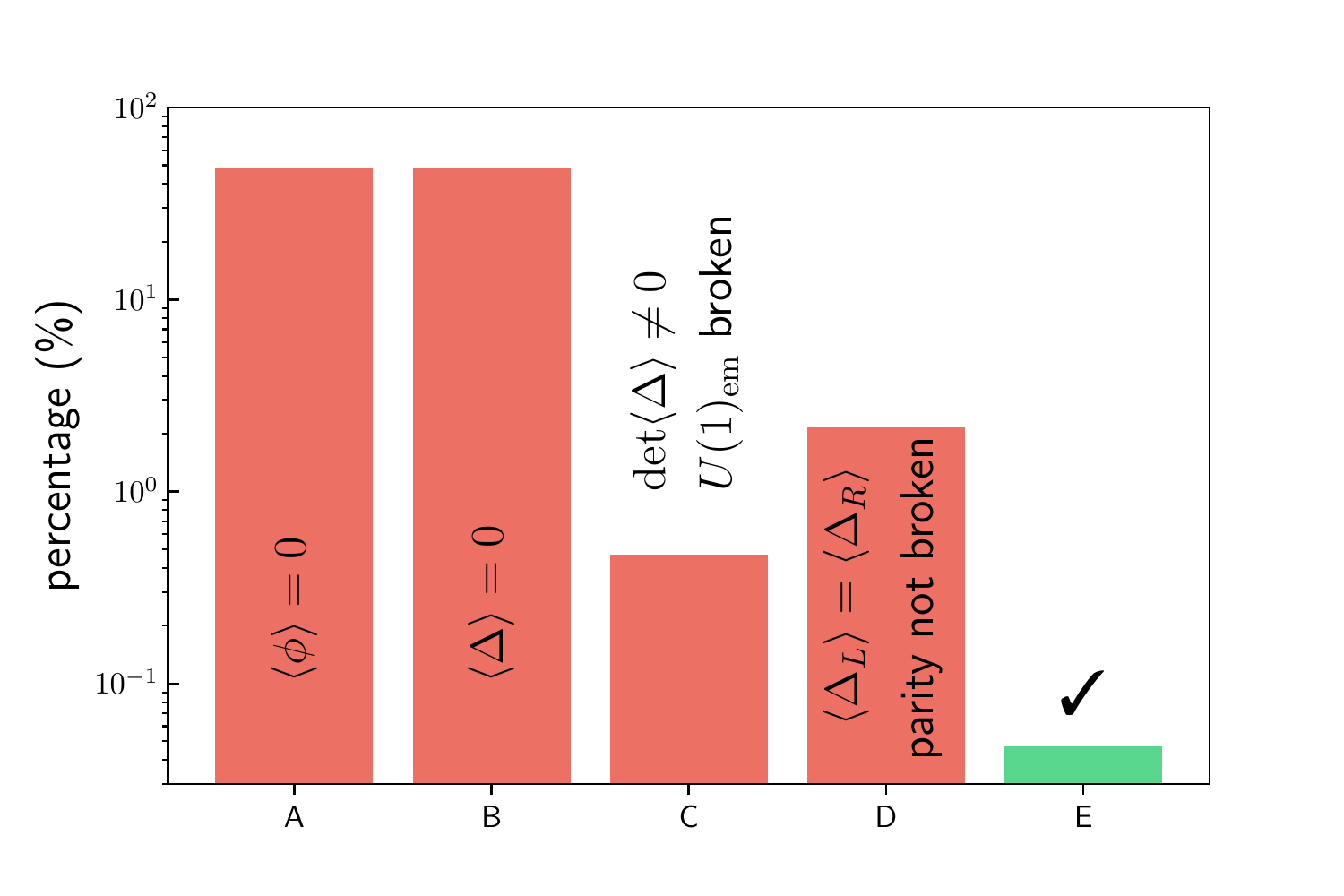}

\caption{\label{fig:bar}Pecentages of the samples leading to the five types
of vacua (type A: 48.6\%, type B: 48.7\%, type C: 0.5\%, type D: 2.2\%,
type E: 0.05\%) classified by the conditions in Eq.~(\ref{eq:LRV-good}).
Only type E corresponds to viable symmetry breaking of LRSM. More
details of the five types of VEV alignments are listed in Tab.~\ref{tab:t}.
To generate the samples, we randomly set the potential parameters
in Eq.~(\ref{eq:LRV-4}), check the BFB condition, and numerically
minimize the potentials.   }
\end{figure}

\begin{table*}
\caption{\label{tab:t}VEV alignments of various types of minima found in the
numerical search. Except for type E, each type is defined by a relation
that violates the good vacuum conditions in Eq.~(\ref{eq:LRV-good}). The typical VEV alignments
in each type are listed in the 3rd--5th columns, with  ``$c_{i}$''
($i=1, 2, 3 , 4$) standing for independent nonzero values. Note that
other VEV alignments that can be converted to the ones in this table
are not shown (see the main text for more details about this issue),
``$\cdots$'' indicates that the enumeration is not exhaustive. 
}

\begin{ruledtabular}
\begin{tabular}{cccccc}
type & definition & $\langle\phi\rangle$ & $\langle\Delta_{L}\rangle$ & $\langle\Delta_{R}\rangle$ & \tabularnewline
 &  &  &  &  & \tabularnewline
\hline 
A &  %
\begin{minipage}[t]{0.2\textwidth}%
violate Eq.~(\ref{eq:LRV-good}) by  $\langle\phi\rangle=0$%
\end{minipage} & $\begin{aligned}\\
\left[\begin{array}{cc}
0 & 0\\
0 & 0
\end{array}\right]
\end{aligned}
$ & $\begin{aligned}\\
\left[\begin{array}{cc}
0 & 0\\
c_{1} & 0
\end{array}\right]
\end{aligned}
$ & $\begin{aligned}\\
\left[\begin{array}{cc}
0 & 0\\
c_{1} & 0
\end{array}\right]
\end{aligned}
$ & \tabularnewline
 &  & $\begin{aligned}\\
\left[\begin{array}{cc}
0 & 0\\
0 & 0
\end{array}\right]
\end{aligned}
$ & $\begin{aligned}\\
\left[\begin{array}{cc}
c_{1} & 0\\
0 & -c_{1}
\end{array}\right]
\end{aligned}
$ & $\begin{aligned}\\
\left[\begin{array}{cc}
c_{1} & 0\\
0 & -c_{1}
\end{array}\right]
\end{aligned}
$ & \tabularnewline
 &  & $\begin{aligned}\\
\left[\begin{array}{cc}
0 & 0\\
0 & 0
\end{array}\right]
\end{aligned}
$ & $\begin{aligned}\\
\left[\begin{array}{cc}
0 & 0\\
0 & 0
\end{array}\right]
\end{aligned}
$ & $\begin{aligned}\\
\left[\begin{array}{cc}
0 & 0\\
0 & 0
\end{array}\right]
\end{aligned}
$ & \tabularnewline
 &  & $\cdots$ & $\cdots$ & $\cdots$ & \tabularnewline
\hline 
B & %
\begin{minipage}[t]{0.2\textwidth}%
violate Eq.~(\ref{eq:LRV-good}) by  $\langle\Delta_L\rangle=\langle\Delta_R\rangle=0$ %
\end{minipage} & $\begin{aligned}\\
\left[\begin{array}{cc}
c_{1} & 0\\
0 & c_{1}
\end{array}\right]
\end{aligned}
$ & $\begin{aligned}\\
\left[\begin{array}{cc}
0 & 0\\
0 & 0
\end{array}\right]
\end{aligned}
$ & $\begin{aligned}\\
\left[\begin{array}{cc}
0 & 0\\
0 & 0
\end{array}\right]
\end{aligned}
$ & \tabularnewline
 &  & $\begin{aligned}\\
\left[\begin{array}{cc}
c_{1} & 0\\
0 & c_{3}
\end{array}\right]
\end{aligned}
$ & $\begin{aligned}\\
\left[\begin{array}{cc}
0 & 0\\
0 & 0
\end{array}\right]
\end{aligned}
$ & $\begin{aligned}\\
\left[\begin{array}{cc}
0 & 0\\
0 & 0
\end{array}\right]
\end{aligned}
$ & \tabularnewline
 &  &  &  &  & \tabularnewline
\hline 
C & %
\begin{minipage}[t]{0.2\textwidth}%
violate Eq.~(\ref{eq:LRV-good}) by  $\det\langle\Delta_{L}\rangle$
or $\det\langle\Delta_{R}\rangle\neq0$ %
\end{minipage} & $\begin{aligned}\\
\left[\begin{array}{cc}
c_{1} & 0\\
0 & c_{1}
\end{array}\right]
\end{aligned}
$ & $\begin{aligned}\\
\left[\begin{array}{cc}
c_{2} & 0\\
0 & -c_{2}
\end{array}\right]
\end{aligned}
$ & $\begin{aligned}\\
\left[\begin{array}{cc}
c_{2} & 0\\
0 & -c_{2}
\end{array}\right]
\end{aligned}
$ & \tabularnewline
 &  & $\begin{aligned}\\
\left[\begin{array}{cc}
c_{1} & 0\\
0 & c_{3}
\end{array}\right]
\end{aligned}
$ & $\begin{aligned}\\
\left[\begin{array}{cc}
c_{2} & 0\\
0 & -c_{2}
\end{array}\right]
\end{aligned}
$ & $\begin{aligned}\\
\left[\begin{array}{cc}
c_{2} & 0\\
0 & -c_{2}
\end{array}\right]
\end{aligned}
$ & \tabularnewline
 &  & $\cdots$ & $\cdots$ & $\cdots$ & \tabularnewline
\hline 
D & %
\begin{minipage}[t]{0.2\textwidth}%
violate Eq.~(\ref{eq:LRV-good}) by  $\langle\Delta_{L}\rangle=\langle\Delta_{R}\rangle$ %
\end{minipage} & $\begin{aligned}\\
\left[\begin{array}{cc}
c_{1} & 0\\
0 & c_{3}
\end{array}\right]
\end{aligned}
$ & $\begin{aligned}\\
\left[\begin{array}{cc}
0 & 0\\
c_{2} & 0
\end{array}\right]
\end{aligned}
$ & $\begin{aligned}\\
\left[\begin{array}{cc}
0 & 0\\
c_{2} & 0
\end{array}\right]
\end{aligned}
$ & \tabularnewline
 &  &  &  &  & \tabularnewline
\hline 
E & satisfy Eq.~(\ref{eq:LRV-good}) & $\begin{aligned}\\
\left[\begin{array}{cc}
c_{1} & 0\\
0 & c_{2}
\end{array}\right]
\end{aligned}
$ & $\begin{aligned}\\
\left[\begin{array}{cc}
0 & 0\\
c_{3} & 0
\end{array}\right]
\end{aligned}
$ & $\begin{aligned}\\
\left[\begin{array}{cc}
0 & 0\\
c_{4} & 0
\end{array}\right]
\end{aligned}
$ & \tabularnewline
 &  &  &  &  & \tabularnewline
\end{tabular}\end{ruledtabular}

\end{table*}

\begin{figure}
\centering

\includegraphics[width=0.5\textwidth]{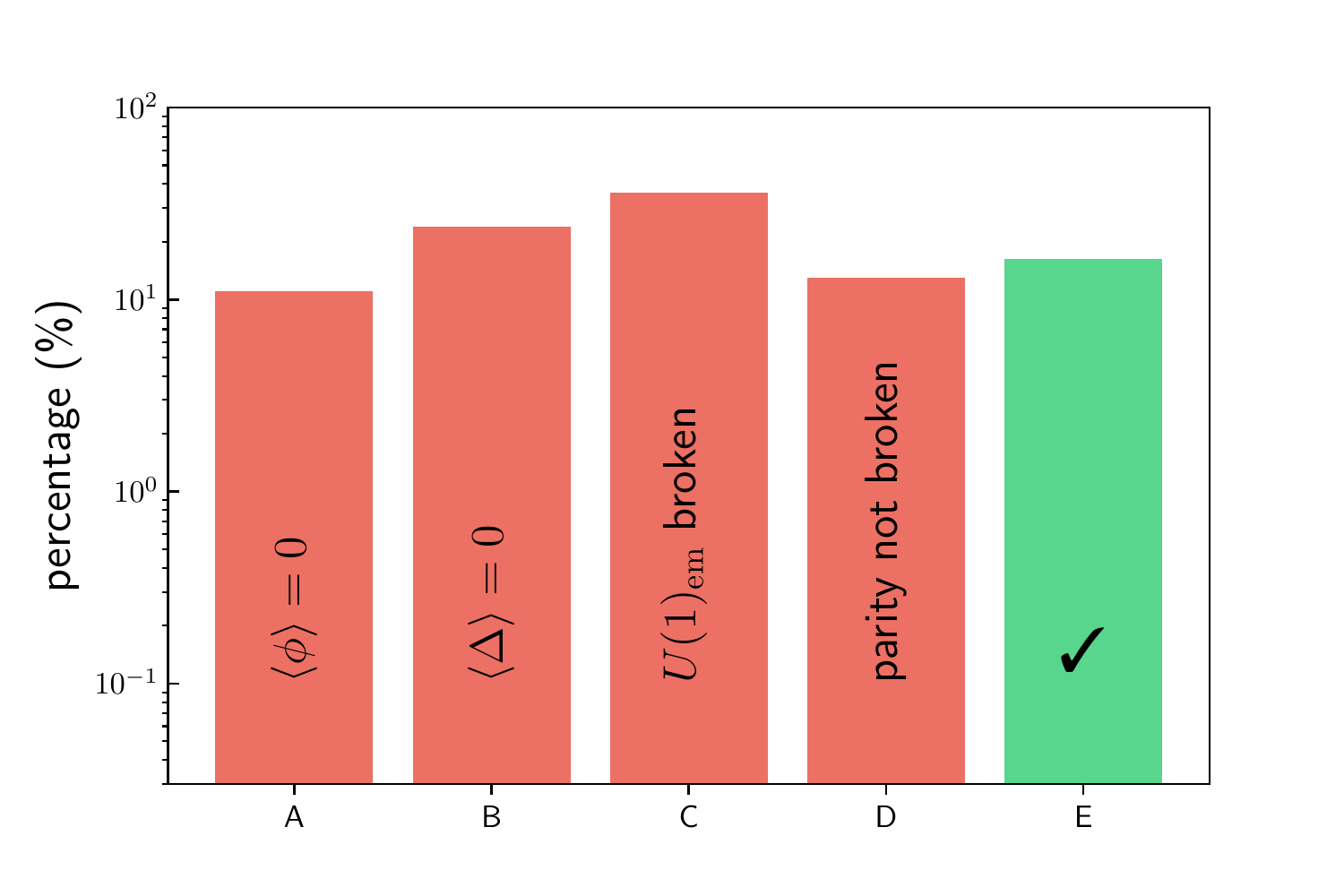}\includegraphics[width=0.5\textwidth]{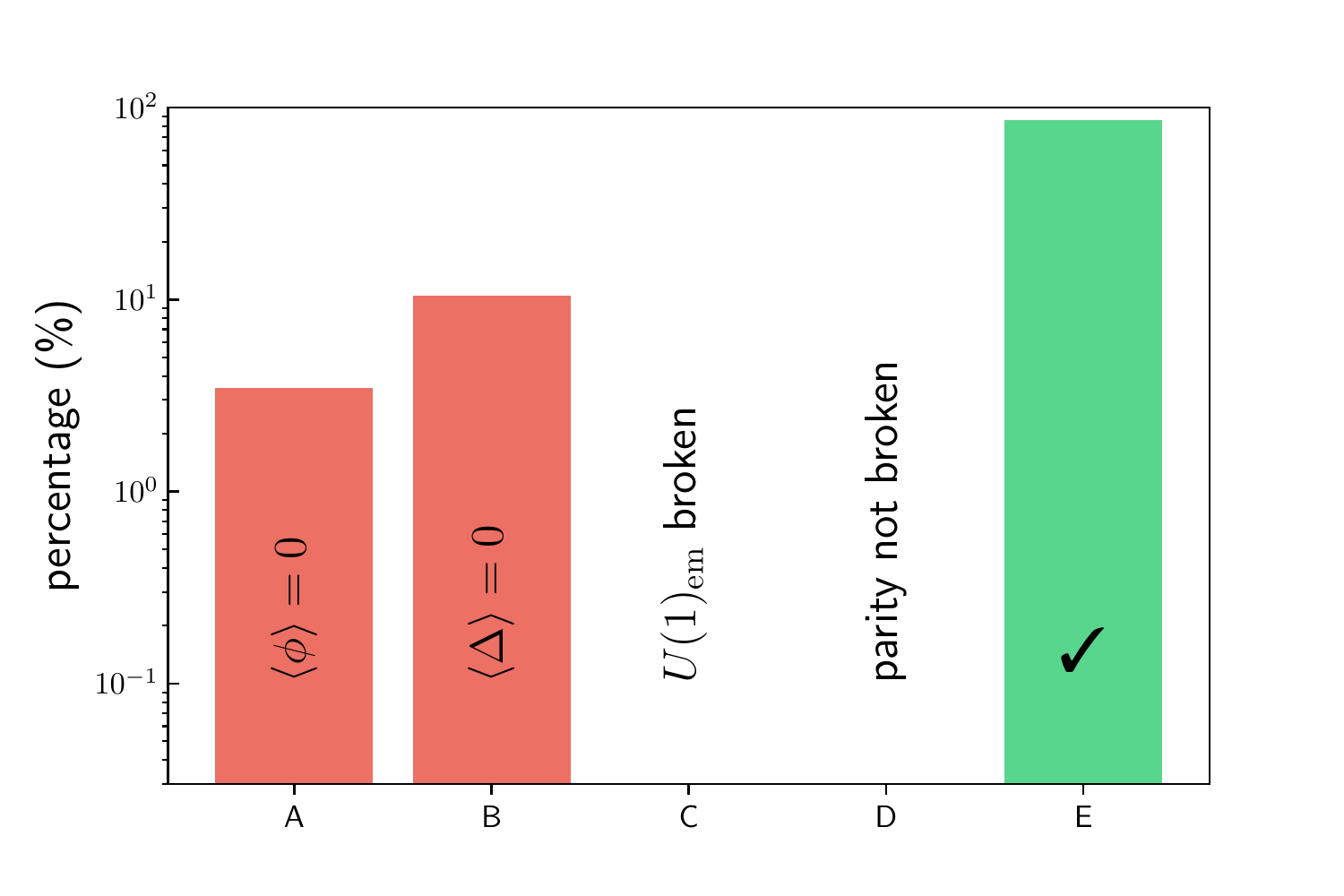}

\caption{\label{fig:bar-1}Similar to Fig.~\ref{fig:bar} except that the
potential parameters are generated with additional constraints given
by Eq.~(\ref{eq:LRV-59}) and Eq.~(\ref{eq:LRV-60}) so that the probability
of obtaining type E vacua is significantly enhanced.  Left: all potential
parameters are positive, $\alpha_{1,\thinspace2,\thinspace3}$ are
suppressed by a small factor and $\beta_{1,\thinspace2,\thinspace3}=0$;
Right: in addition to the constraints used in the left panel, we require that $\lambda_{2}=\lambda_{4}=\rho_{4}=0$,
$\rho_{3}>2\rho_{1}$. The percentages from left to right are 11.0\%
(A), 23.9\% (B), 35.9\% (C), 13.0\% (D), 16.2\% (E); 3.4\% (A), 10.5\%
(B), 0 (C), 0 (D), 86.0\% (E).}
\end{figure}

We apply the above numerical method to the LRSM potential in Eq.~(\ref{eq:LRV-4}).
First, let us arbitrarily set the values of potential parameters to
see in general what VEV alignments would be obtained.   Both the
quadratic and quartic couplings are generated by a uniform distribution
in the interval $[-4\pi,\ 4\pi]$. In this work, the energy scales
of all dimensional quantities such as the quadratic couplings and
the field values are not relevant. For example, we can use 
$v\equiv246$ GeV as the energy unit, instead of GeV or TeV, then $\mu_{1}^{2}=0.5\times\left(246\,\,{\rm GeV}\right)^{2}=0.5v^{2}$
can be simply written as $0.5$ in the computer program. 

The quartic couplings generated in the above way can not guarantee
that the potential is bounded from below. The BFB check can
be numerically performed by setting the quadratic couplings to zero
and then run the minimization process. If any point with $V<0$ is
reached during minimization, then the potential is not BFB and the
sample is abandoned. If a sample passes the BFB check, then we further
minimize the potential with the nonzero quadratic couplings. 

When a minimum is successfully obtained  in this process, we check
if it violates the four good vacuum conditions in Eq.~(\ref{eq:LRV-good}), from
(a) to (d) sequentially. If any of them is violated, then we stop
checking the remaining conditions and tag it as type A, B, C, or D,
corresponding to the violation of condition (a), (b), (c), or (d), 
respectively. If all the conditions are satisfied, it is tagged as
type E, a good and successful vacuum. 

In Fig.~\ref{fig:bar}, we present the result of the above analysis
on 144891 samples (all passing the BFB check). In this randomly generated
data set, 70380, 70625, 683, and 3135 of the samples fall into the
categories of type A, B, C, and D, respectively.  Only 68 samples
are of type E, which is about 0.05\% of the total number. We further
inspect the VEV alignments of all samples of the five types and
list them in  in Tab.~\ref{tab:t}. As we have discussed in Sec.~\ref{sub:bad-vacuum},
some symmetry transformations can transform the VEV alignments from
one form to another---see, e.g. Eq.~(\ref{eq:LRV-58}). For such cases,
in Tab.~\ref{tab:t} we only list the representative forms. More
specifically, whenever appropriate $U_{L}$ and $U_{R}$ transformations are
allowed (e.g.\ $\langle\phi\rangle=0$, or $\kappa_{1}=\kappa_{2}$,
or $U_{L}=U_{R}=i\sigma_{2}$), we always use them to transform $\Delta_{L}$
and $\Delta_{R}$ to lower triangular forms according to the Schur
decomposition. 

The low percentage (0.05\%) of type E can be understood from the
analytical studies in Sec.~\ref{sec:analytical}. First, for those
simplified cases we have studied, one can see that the quadratic couplings
$\mu_{1}^{2}$, $\mu_{2}^{2}$ , and $\mu_{3}^{2}$ have to be positive
to get nonzero $\langle\phi\rangle$ and $\langle\Delta_{L/R}\rangle$.
Let us assume that for more general potentials (e.g.\ $\lambda_{4}$,
$\beta_{1,\thinspace2,\thinspace3}$ are no longer zero) this conclusion
approximately holds as well. Then requiring the three quadratic couplings to
be positive in the random number generation already produces a factor
of $(1/2)^{3}=1/8$ which suppresses the percentage by one order of
magnitude. Moreover, in Eq.~(\ref{eq:LRV-44}) and Eq.~(\ref{eq:LRV-45})
some quartic couplings may also need to be positive to get a good
vacuum. If 11 of the 17 parameters in the full potential are required
to be positive, the suppression factor can easily reach $(1/2)^{11}\approx0.05\%$.
Some parameters may contribute suppression factors smaller or larger
than $1/2$, say $1/p$. Generally it is possible to get a significant
suppression at the order of $(1/p)^{n}$ where $n\leq17$. This explains
why the percentage can be suppressed to the level of 0.05\%.

Although the suppression is understandable, it is would be better
to avoid the suppression or at least to know a part of the parameter
space that would lead to the correct symmetry breaking with a much
higher probability. According to our analytical study, we are led to simple conditions to enhance the probability of ending in a good vacuum: 
\begin{equation}
\begin{cases}
0\leq\mu_{i}^{2}\leq4\pi v^{2} & (i=1,\thinspace2,\thinspace3),\\
0\leq\lambda_{i},\ \rho_{i}\leq4\pi & (i=1,\thinspace2,\thinspace3,\thinspace4),\\
0\leq\alpha_{i}\leq0.2\pi & (i=1,\thinspace2,\thinspace3),\\
\beta_{i}=0 & (i=1,\thinspace2,\thinspace3).
\end{cases}\ \label{eq:LRV-59}
\end{equation}
Requiring the constraints in Eq.~(\ref{eq:LRV-59}), we repeat the numerical
process used to generate Fig.~\ref{fig:bar} and obtain the left
plot in Fig.~\ref{fig:bar-1}. As the plot shows, with these constraints,
the percentage of type E is enhanced to 16.2\%, which is at the same
order of magnitude as the other types. Eq.~(\ref{eq:LRV-59}) is proposed
based on the analytical result summarized in Eqs.~(\ref{eq:LRV-43})-(\ref{eq:LRV-46}),
but it  allows more general parameter settings, e.g.\ $\lambda_{2}$,
$\rho_{2,\thinspace4}$, $\alpha_{1,\thinspace2,\thinspace3}$ do
not have to be fixed to zero. It is a compromise between the generality
(also simplicity) and the enhancement of the percentage.

Including more constraints from Eqs.~(\ref{eq:LRV-43})-(\ref{eq:LRV-46})
can further enhance the percentage at the cost of loss of generality.
In the right panel of Fig.~\ref{fig:bar-1}, we include the constraints
\begin{equation}
\lambda_{2}=\lambda_{4}=\rho_{4}=0,\ \rho_{3}>2\rho_{1},\label{eq:LRV-60}
\end{equation}
together with Eq.~(\ref{eq:LRV-59}) and obtain a much higher percentage
(86\%) of type E, which means the majority of the samples generated
under these constraints have type E vacua.

\begin{figure}
\centering

\includegraphics[width=0.38\textwidth]{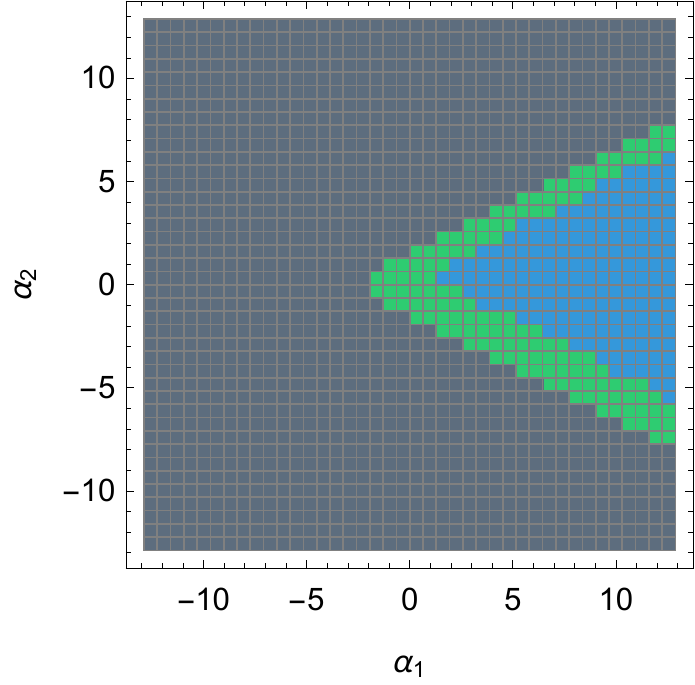}\ \includegraphics[width=0.38\textwidth]{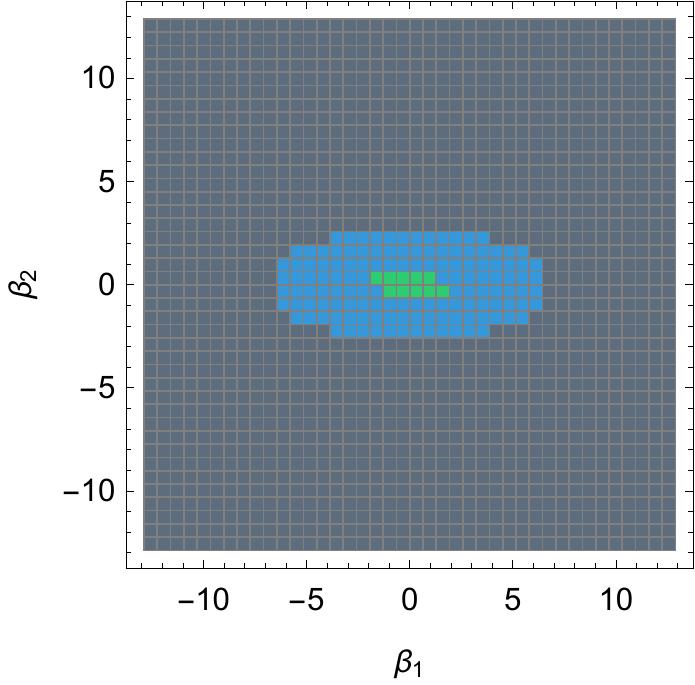}

\vspace{0.3cm}

\includegraphics[width=0.38\textwidth]{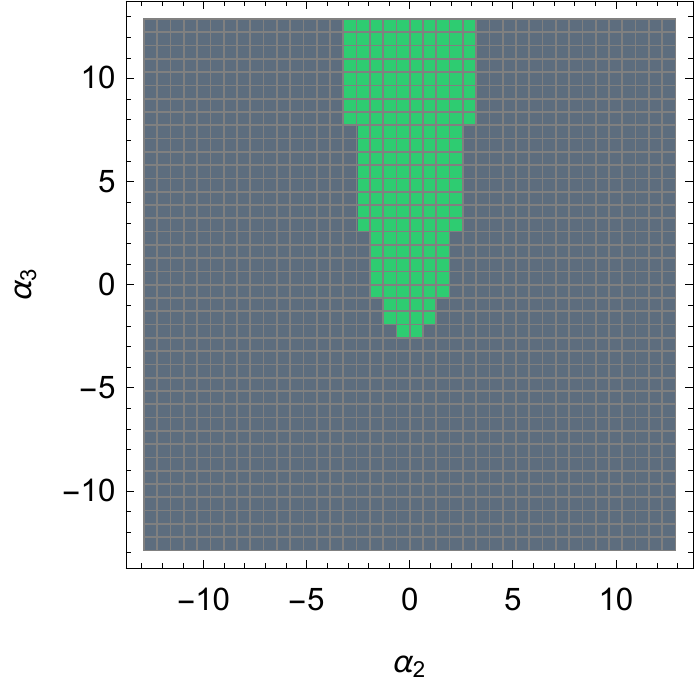}\ \includegraphics[width=0.38\textwidth]{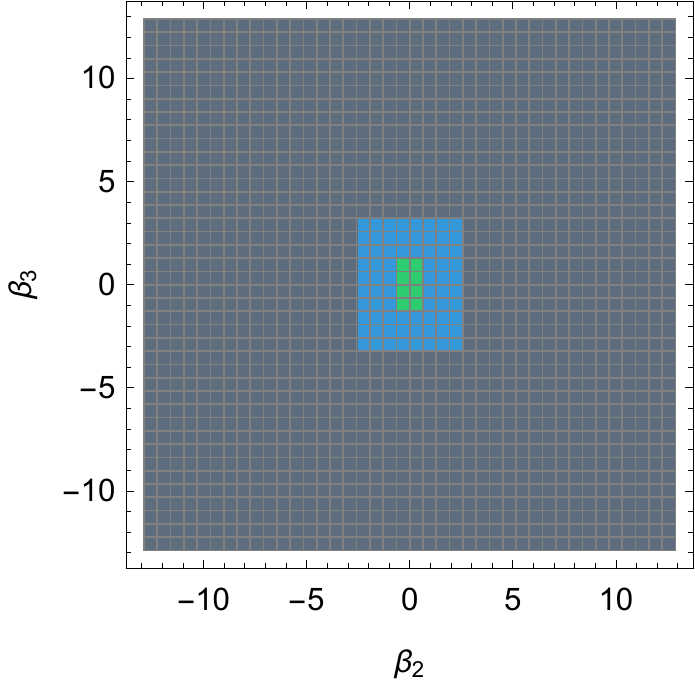}

\vspace{0.3cm}

\includegraphics[width=0.38\textwidth]{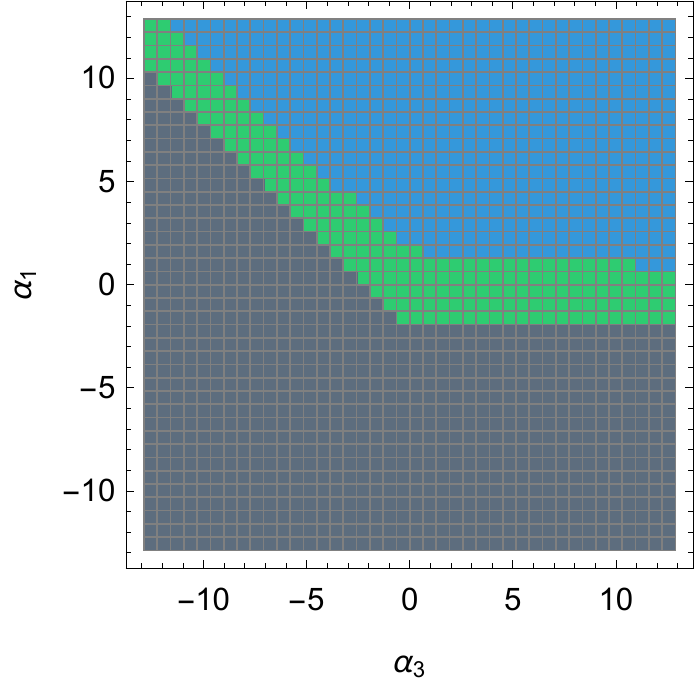}\ \includegraphics[width=0.38\textwidth]{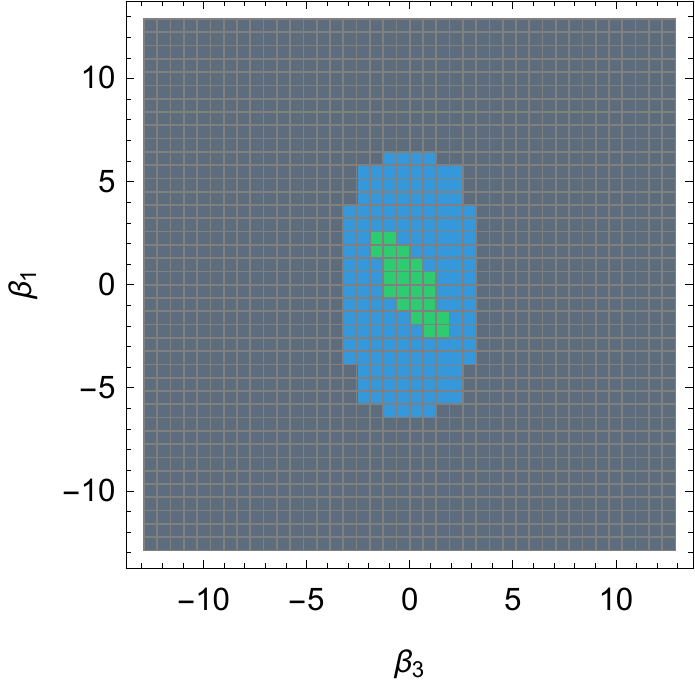}\caption{\label{fig:a123b123} Global minimum constraints on $\alpha$ and
$\beta$. Samples in the green region have type E global minima so
they will lead to successful symmetry breaking; the blue region violates
the global minimum constraints which means that either the potential
does not have type E minimum or its type E minimum is local; the black
region violates the BFB condition. Other potential parameters, if
not indicated by the plots, are fixed at the benchmark values in Eq.~(\ref{eq:LRV-63}).
The grid interval is $0.2\pi$.}
\end{figure}

\vspace{0.3cm}

As we have seen that among those randomly generated samples some may
have type E vacua and some may not, if the potential is required
to lead to successful symmetry breaking, it must be subjected to a
lot of constraints. Below we would like to study such constraints. 

Note that these constraints are not fully equivalent to the requirement
that the potential has a type E minimum. If the potential has no type
E minimum, then of course it can not lead to successful symmetry breaking.
But even if it has a type E minimum, the minimum could be a local
one which coexists with other much deeper minima. Then the vacuum
at the type E minimum is not absolutely stable as it may decay to
other deeper vacua via quantum tunneling or thermal fluctuation. There
is a possibility that the decay rate is very low so that the lifetime
is longer than the age of the Universe, known as the meta-stability.
Since the analyses including meta-stability would be too much involved here,
for simplicity, we only consider the absolute stability. Therefore
in what follows, when we claim that a potential can lead to successful
symmetry breaking, we mean the potential has a global minimum of type
E. The corresponding constraints will be referred to as \emph{the
global minimum constraints}.

Let us investigate the effect of global vs.\ local minimum. Since there
is no minimization algorithm that can guarantee to find global minima
for general cases, the method of global minimum test used in our program
is repeating the minimization process with random initial values for
many times. If none of these minima is deeper than the one being tested,
the more likely it is a global minimum. Obviously
the more times the process is repeated, the more likely it is
a global minimum. This method still can not guarantee the correctness of the global minimum, 
but with a large number of repetitions, the result will be very reliable. 
To quantify the effect of testing for a global minimum, we present now constraints on some 
parameters as illustration, choosing the six $\alpha_{1,\thinspace2,\thinspace3}$
and $\beta_{1,\thinspace2,\thinspace3}$ parameters. 
In Fig.~\ref{fig:a123b123} the green region has 
successful symmetry breaking with a global minimum, while the blue region only is a global minimum. 
Each plot shows the constraint on a pair
of the parameters ($\alpha_{i}-\alpha_{j}$ or $\beta_{i}-\beta_{j}$)
while the other parameters are fixed at:
\begin{equation}
{\rm benchmark:\ }\begin{cases}
(\mu_{1}^{2},\ \mu_{2}^{2},\ \mu_{3}^{2}) & =(5.0,\ 2.0,\ 9.0)v^{2}\,,\\
(\lambda_{1},\ \lambda_{2},\ \lambda_{3},\ \lambda_{4}) & =(1.0,\ 0.0,\ 1.2,\ 0.0)\,,\\
(\rho_{1},\ \rho_{2},\ \rho_{3},\ \rho_{4}) & =(1.0,\ 0.2,\ 3.0,\ 0.0)\,,\\
(\alpha_{1},\ \alpha_{2},\ \alpha_{3}) & =(0.5,\ 0.0,\ 0.7)\,,\\
(\beta_{1},\ \beta_{2},\ \beta_{3}) & =(0.0,\ 0.0,\ 0.0)\,.
\end{cases}\ \label{eq:LRV-63}
\end{equation}
This benchmark is set in such a way that by default (i.e.\ no parameters
are changed) it has a global minimum of type E. The plots are produced
in coarse grids because for each sample the program has to run the
global minimum test for many times which is CPU intensive.  So currently
we cannot compute too many  samples with limited computer power and
consequently the interval of grid scan cannot be too small. In Fig.~\ref{fig:a123b123}
we use $40\times40$ grids in the range $[-4\pi,\ 4\pi]^{2}$ with
a interval of $0.2\pi$.

As one can see in the left panels of Fig.~\ref{fig:a123b123}, the
green regions cover the central point $\alpha_{1,\thinspace2,\thinspace3}=0$
and the nearby part (within 3 or 4 blocks) is also green. This implies
that small $\alpha_i$ indeed can lead to absolutely stable type E
vacua, which is a conjecture of our analytical study. This is
approximately true also for the $\beta$ parameters. However, the difference
is that the $\alpha_i$ do not have to be small (in some direction they 
can reach $4\pi$) while the $\beta_i$, at least for this benchmark, have
to be small.

We also note that Fig.~3 and Fig.~4 establish our claim in the introduction that the BFB conditions only provide a necessary but not sufficient condition for an acceptable vacuum since all the columns in Fig.~3 and Fig.~4 satisfy the BFB condition whereas only the green column satisfies the desired vacuum condition.

\begin{figure}
\centering

\includegraphics[width=9cm]{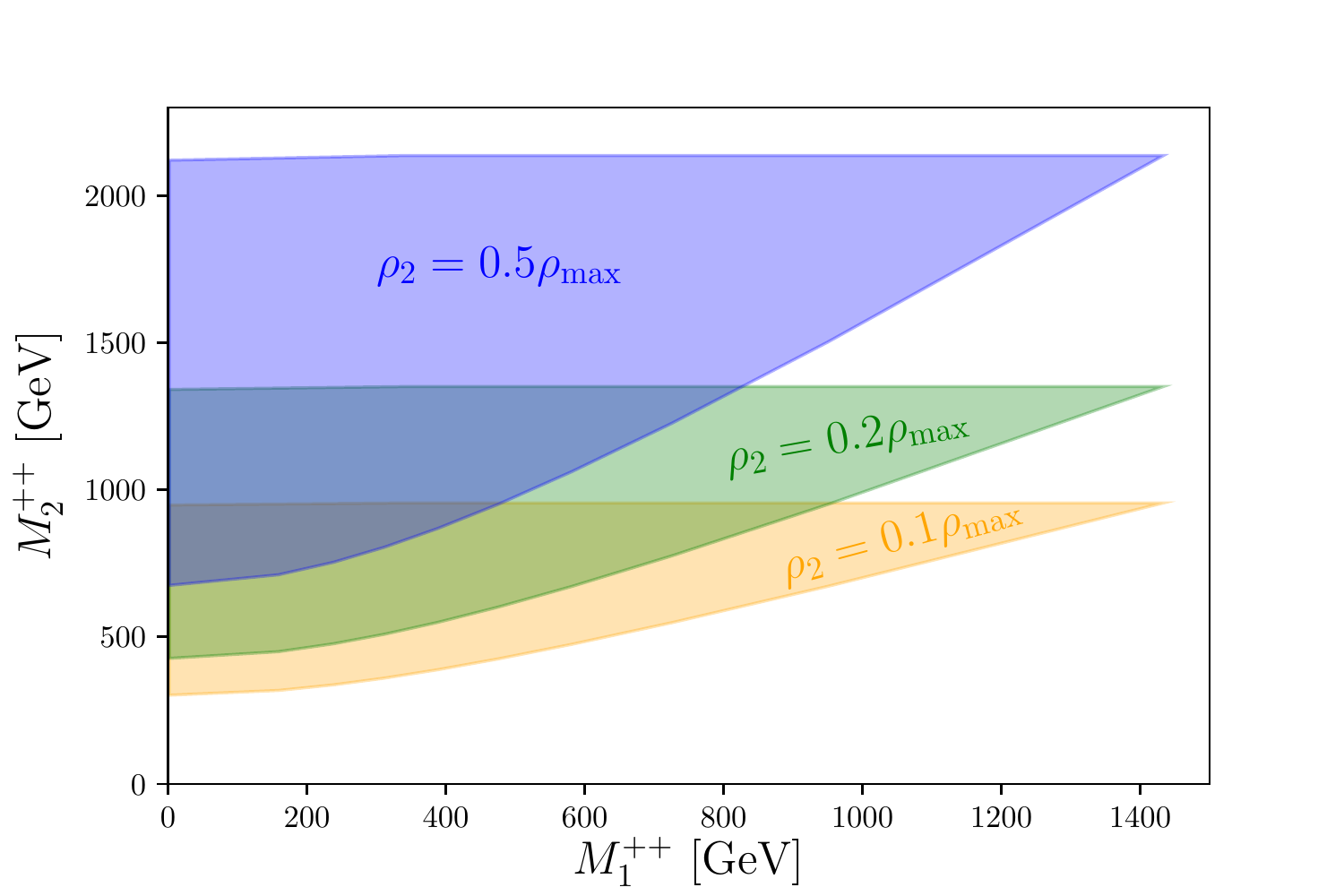}

\caption{\label{fig:mass-constraint}The global minimum constraint on the two doubly 
charged scalar masses $(M_{1}^{\pm\pm},\ M_{2}^{\pm\pm})$ for the
example given by Eq.~(\ref{eq:para-set1}).}
\end{figure}

Because the scalar mass spectrum is fully determined by the parameters
of the scalar potential, the global minimum constraints on the potential
parameters can be converted to constraints on the scalar mass spectrum.
After symmetry breaking the scalar sector contains
(including the Goldstone bosons) eight electrically neutral bosons,
four singly charged bosons and two doubly charged bosons, among which
the bosons with the same electric charge generally have mass mixing.
Therefore the mass matrices of the neutral and singly charged scalar
bosons are quite complicated, but the mass matrix of doubly charged
bosons is much simpler. 
For simplicity, we will thus only discuss the mass spectrum of the doubly
charged bosons. Their mass matrix  is
\begin{equation}
{\cal M}_{11}^{\pm\pm}=\left(\begin{array}{cc}
{\cal M}_{11}^{\pm\pm} & {\cal M}_{12}^{\pm\pm}\\
{\cal M}_{21}^{\pm\pm} & {\cal M}_{22}^{\pm\pm}
\end{array}\right),\label{eq:mass-doubly-ch}
\end{equation}
where
\[
{\cal M}_{11}^{\pm\pm}=\frac{1}{2}(\rho_{3}-2\rho_{1})v_{R}^{2}+2\rho_{2}v_{L}^{2}+\frac{1}{2}\alpha_{3}(\kappa_{1}^{2}-\kappa_{2}^{2})\,,
\]
\[
{\cal M}_{22}^{\pm\pm}=2\rho_{2}v_{R}^{2}+\frac{1}{2}(\rho_{3}-2\rho_{1})v_{L}^{2}+\frac{1}{2}\alpha_{3}(\kappa_{1}^{2}-\kappa_{2}^{2})\,,
\]
\[
{\cal M}_{12}^{\pm\pm}=\left({\cal M}_{21}^{\pm\pm}\right)^{*}=2\rho_{4}v_{R}v_{L}e^{-i\theta_{L}}+\frac{1}{2}\left(\beta_{1}\kappa_{1}\kappa_{2}e^{-i\theta_{2}}+\beta_{2}\kappa_{2}^{2}e^{-2i\theta_{2}}+\beta_{3}\kappa_{1}^{2}\right).
\]
Note that if $\rho_{4}$ and $\beta_{1,\thinspace2,\thinspace3}$
are set to zero, then the mass matrix is diagonal and one can immediately
obtain the eigenvalues (i.e.\ the mass squares of the two doubly charged
bosons):
\begin{equation}
\left(M_{1}^{\pm\pm}\right)^{2}=(\rho_{3}-2\rho_{1})v_{R}^{2}/2+\alpha_{3}(\kappa_{1}^{2}-\kappa_{2}^{2})/2\,, \ \  \left(M_{2}^{\pm\pm}\right)^{2}=2\rho_{2}v_{R}^{2}+\alpha_{3}(\kappa_{1}^{2}-\kappa_{2}^{2})/2\,. 
\end{equation}
This greatly simplifies the scenario and we would like to take it
as an  example to show the global minimum constraints on the mass
spectrum. We also set other parameters to the following specific values:
\begin{equation}
\begin{cases}
(\mu_{1}^{2},\ \mu_{2}^{2},\ \mu_{3}^{2}) & =(0.3,\ 0.2,\ 9)v_{X}^{2},\\
(\lambda_{1},\ \lambda_{2},\ \lambda_{3},\ \lambda_{4}) & =(0.13,\ 0,\ 0.6,\ 0),\\
(\alpha_{1},\ \alpha_{2},\ \alpha_{3}) & =(0,\ 0,\ 1\times10^{-4}),
\end{cases}\ 
\label{eq:para-set1}
\end{equation}
except for $\rho_{1}$, $\rho_{2}$ and $\rho_{3}$. We take $\rho_{1}$ and $\rho_{3}$ as free parameters ranging from $-\rho_{\max}$ to $\rho_{\max}$, and fix $\rho_{2}$ at some values indicated in Fig.~\ref{fig:mass-constraint}. Here $\rho_{\max}$ is set at a small value $10^{-3}$ (for a reason to be explained below) and $v_{X}^{2}$ is a floating energy scale which is always tuned
to make  $\sqrt{\kappa_{1}^{2}+\kappa_{2}^{2}}=246$ GeV. 

To make the scenario considered here more realistic, we also require that it contains a SM-like Higgs boson with the mass $m_h\approx 125$ GeV and a large $v_R$ so that the mass of $W_R$ is above the LHC constraints. The parameters in Eq.~(\ref{eq:para-set1}) have been tuned in such a way that for $\rho_{1}$ and $\rho_{3}$ varying in  $[-\rho_{\max},\, \rho_{\max}]$, the SM Higgs mass $m_h$ ranges within $125\pm 1.5$ GeV and $v_R$ ranges from 21 to 67 TeV.
We have checked that changing  $\rho_{\max}$ within one order of magnitude leads to very little change of $(M_{1}^{\pm\pm},\ M_{2}^{\pm\pm})$ or  $m_h$. Changing  $\rho_{\max}$, however, has significant impact on $v_R$. Generally larger   $\rho_{\max}$ leads to smaller $v_R$, which is the reason why we use a small $\rho_{\max}$ here.

With the above parameter setting, we scan the parameter space and compute the mass spectrum of the doubly charged Higgs bosons when the sample satisfies the global minimum requirement. The result is shown in Fig.~\ref{fig:mass-constraint}, where the yellow, green, and blue regions are the allowed regions by the global minimum requirement for $\rho_{2}=$ $0.1 \rho_{\max}$, $0.2\rho_{\max}$, and $0.5\rho_{\max}$ respectively.
Note that the constraints  presented in Fig.~\ref{fig:mass-constraint} are only for a very specific parameter setting so they should not be interpreted as universal constraints on the mass spectra. 
Changing the parameter setting can easily lead to significant changes of the constraints, as illustrated by different values of $\rho_2$ in Fig.~\ref{fig:mass-constraint}. It is still interesting that there are certain mass ranges for the doubly charged scalars which are forbidden by our analysis and can be used to test the model.

\section{Conclusion\label{sec:Conclusion}}
\noindent 
We have performed a study on the vacuum structure of the
left-right symmetric scalar potential. 
The goal was to investigate whether the usually considered VEV alignment in Eq.~(\ref{eq:LRV-1}) can be obtained from a generic scalar potential as a global minimum. General criteria to identify a charge conserving and parity violating vacuum were derived (see Eq.\  (\ref{eq:LRV-good}), and it was demonstrated that the potential 
parameters are subject to many constraints in order to achieve this minimum. In general if we do
not put any constraints on the potential parameters, 
as indicated by Fig.~\ref{fig:bar}, the probability to end up within the desired 
VEV alignment is very low, only 0.05\%. 

We have also analytically studied the minima of the potential in the
absence of some terms and obtained conditions that enable us
to obtain the correct VEV alignment more easily, as shown
in Fig.~\ref{fig:bar-1}. By requiring that the corresponding minimum
is global in this case, we also illustrate the constraints on the potential parameters
in Fig.~\ref{fig:a123b123}.

Our work suggests that successful generation of the usually considered
VEV alignment in standard left-right symmetric theories and keeping the vacuum absolutely stable would
produce important constraints on the parameters of the potential. These constraints
may have interesting phenomenological consequences such as constraints
on the mass spectrum of scalar bosons, or the Higgs self-couplings,
etc., which can be used to test the model. We have given the example of doubly charged scalar masses in the model as an example.

The present paper can be a starting point for further and much more involved analyses along these lines, such as analyzing loop-corrected effective potentials, 
investigating the vacuum lifetime of non-global minima,  more phenomenological consequences of the global minima, or studies of alternative left-right symmetric models.

\begin{acknowledgments}
\noindent We thank Werner Porod for clarifying comments. 
The work of B.D.\ is supported by the US Department of Energy under Grant No.\ DE-SC0017987, R.N.M.\ is supported by the US National Science Foundation under Grant No.\ PHY1620074, and 
W.R.\ is supported by the DFG with grant RO 2516/7-1 in the Heisenberg program. 

\end{acknowledgments}

\appendix

\section{Deriviation of the seesaw relation of VEVs\label{sec:seesaw}}
\noindent
In this appendix, we review the derivation of the seesaw relation between the left and right $\Delta$ vevs. To do this let us
first, let us compute the six derivatives in Eq.~(\ref{eq:LRV-6})
explicitly: 
\begin{eqnarray}
0=\frac{\partial V}{\partial\kappa_{1}} & = & 8\kappa_{2}^{2}\kappa_{1}\lambda_{2}\cos2\theta_{2}+2\kappa_{1}^{3}\lambda_{1}+2\kappa_{2}^{2}\kappa_{1}\lambda_{1}+4\kappa_{2}^{2}\kappa_{1}\lambda_{3}-2\kappa_{1}\mu_{1}^{2}\nonumber \\
 &  & +2\kappa_{2}\cos \theta_{2}\left(3\kappa_{1}^{2}\lambda_{4}+\kappa_{2}^{2}\lambda_{4}-2\mu_{2}^{2}+\alpha_{2}v_{L}^{2}+\alpha_{2}v_{R}^{2}\right) \\\nonumber
 &  & +2\beta_{2}\kappa_{1}v_{L}v_{R}\cos \theta_{L}+\beta_{1}\kappa_{2}v_{L}v_{R}\cos\left(\theta_{2}-\theta_{L}\right)+\alpha_{1}\kappa_{1}v_{L}^{2}+\alpha_{1}\kappa_{1}v_{R}^{2}\,,\label{eq:LRV-48}\\
\nonumber \\
0=\frac{\partial V}{\partial\kappa_{2}} & = & 8\kappa_{1}^{2}\kappa_{2}\lambda_{2}\cos 2\theta_{2} +2\kappa_{2}^{3}\lambda_{1}+2\kappa_{1}^{2}\kappa_{2}\lambda_{1}+4\kappa_{1}^{2}\kappa_{2}\lambda_{3}-2\kappa_{2}\mu_{1}^{2}\nonumber \\
 &  & +2\kappa_{1}\cos \theta_{2}\left(\kappa_{1}^{2}\lambda_{4}+3\kappa_{2}^{2}\lambda_{4}-2\mu_{2}^{2}+\alpha_{2}v_{L}^{2}+\alpha_{2}v_{R}^{2}\right) \\
 &  & +2\beta_{3}\kappa_{2}v_{L}v_{R}\cos\left(2\theta_{2}-\theta_{L}\right)+\beta_{1}\kappa_{1}v_{L}v_{R}\cos\left(\theta_{2}-\theta_{L}\right)\nonumber \\ \nonumber
 &  & +\alpha_{1}\kappa_{2}v_{L}^{2}+\alpha_{3}\kappa_{2}v_{L}^{2}+\alpha_{1}\kappa_{2}v_{R}^{2}+\alpha_{3}\kappa_{2}v_{R}^{2}\,,\label{eq:LRV-49}\\
\nonumber \\
0=\frac{\partial V}{\partial v_{R}} & = & \rho_{3}v_{L}^{2}v_{R}+\beta_{2}\kappa_{1}^{2}v_{L}\cos \theta_{L}+\beta_{3}\kappa_{2}^{2}v_{L}\cos\left(2\theta_{2}-\theta_{L}\right)+\beta_{1}\kappa_{1}\kappa_{2}v_{L}\cos\left(\theta_{2}-\theta_{L}\right)\nonumber \\
 &  & +4\alpha_{2}\kappa_{1}\kappa_{2} v_R \cos \theta_{2}+\alpha_{1}\kappa_{1}^{2}v_{R}+\alpha_{1}\kappa_{2}^{2}v_{R}+\alpha_{3}\kappa_{2}^{2}v_{R}-2\mu_{3}^{2}v_{R}+2\rho_{1}v_{R}^{3}\,,\label{eq:LRV-50}\\
\nonumber \\
0=\frac{\partial V}{\partial v_{L}} & = & \beta_{2}\kappa_{1}^{2}v_{R}\cos \theta_{L}+\beta_{3}\kappa_{2}^{2}v_{R}\cos\left(2\theta_{2}-\theta_{L}\right)+\beta_{1}\kappa_{1}\kappa_{2}v_{R}\cos\left(\theta_{2}-\theta_{L}\right)+\rho_{3}v_{L}v_{R}^{2}\nonumber \\
 &  & +4\alpha_{2}\kappa_{1}\kappa_{2}v_{L} \cos \theta_{2}+\alpha_{1}\kappa_{1}^{2}v_{L}+\alpha_{1}\kappa_{2}^{2}v_{L}+\alpha_{3}\kappa_{2}^{2}v_{L}-2\mu_{3}^{2}v_{L}+2\rho_{1}v_{L}^{3}\,,\label{eq:LRV-51}\\
\nonumber \\
0=\frac{\partial V}{\partial\theta_{2}} & = & -8\kappa_{2}^{2}\kappa_{1}^{2}\lambda_{2}\sin2\theta_{2}-2\kappa_{1}\kappa_{2}\sin\theta_{2}\left(\kappa_{1}^{2}\lambda_{4}+\kappa_{2}^{2}\lambda_{4}-2\mu_{2}^{2}+\alpha_{2}v_{L}^{2}+\alpha_{2}v_{R}^{2}\right)\nonumber \\
 &  & -v_{L}v_{R}\kappa_{2}\left(\beta_{1}\kappa_{1}\sin\left(\theta_{2}-\theta_{L}\right)+2\beta_{3}\kappa_{2}\sin\left(2\theta_{2}-\theta_{L}\right)\right),\label{eq:LRV-52}\\
\nonumber \\
0=\frac{\partial V}{\partial\theta_{L}} & = & -  \beta_{2}\kappa_{1}^{2}v_{L}v_{R} \sin\theta_{L}+\beta_{1}\kappa_{2}\kappa_{1}v_{L}v_{R}\sin\left(\theta_{2}-\theta_{L}\right)+\beta_{3}\kappa_{2}^{2}v_{L}v_{R}\sin\left(2\theta_{2}-\theta_{L}\right).\label{eq:LRV-53}\\
\nonumber 
\end{eqnarray}
The first three equations can be regarded as linear equations of $\mu_{1}^{2}$,
$\mu_{2}^{2}$, and $\mu_{3}^{2}$ so we can solve them with respect
to $\mu_{1}^{2}$, $\mu_{2}^{2}$, and $\mu_{3}^{2}$ without much
effort: 
\begin{eqnarray}
\mu_{1}^{2} & = & \frac{\left(\alpha_{1}\kappa_{1}^{2}-\alpha_{1}\kappa_{2}^{2}-\alpha_{3}\kappa_{2}^{2}\right)\left(v_{L}^{2}+v_{R}^{2}\right)}{2\left(\kappa_{1}^{2}-\kappa_{2}^{2}\right)}+\frac{v_{L}v_{R}\left(\beta_{2}\kappa_{1}^{2}\cos \theta_{L}-\beta_{3}\kappa_{2}^{2}\cos\left(2\theta_{2}-\theta_{L}\right)\right)}{\kappa_{1}^{2}-\kappa_{2}^{2}}\nonumber \\
 &  & +2\kappa_{2}\kappa_{1}\lambda_{4}\cos \theta_{2}+\kappa_{1}^{2}\lambda_{1}+\kappa_{2}^{2}\lambda_{1}\,,\label{eq:LRV-55}\\
\nonumber \\
\mu_{2}^{2} & = & \kappa_{1}\kappa_{2}\sec\theta_{2}\left(2\lambda_{2}\cos2\theta_{2}+\lambda_{3}\right)+\frac{1}{2}\left(\kappa_{1}^{2}+\kappa_{2}^{2}\right)\lambda_{4}+\left(v_{L}^{2}+v_{R}^{2}\right)\left(\frac{\alpha_{3}\kappa_{1}\kappa_{2}\sec\theta_{2}}{4\left(\kappa_{1}^{2}-\kappa_{2}^{2}\right)}+\frac{\alpha_{2}}{2}\right)\nonumber \\
 &  & +\frac{\kappa_{1}\kappa_{2} v_{L}v_{R}\left[\beta_{3}\cos\left(2\theta_{2}-\theta_{L}\right)-\beta_{2}\cos \theta_{L}\right]\sec\theta_{2}}{2\left(\kappa_{1}^{2}-\kappa_{2}^{2}\right)}+\frac{1}{4}\beta_{1}v_{L}v_{R}\sec\theta_{2} \cos\left(\theta_{2}-\theta_{L}\right),\label{eq:LRV-56}\\
\nonumber \\
\mu_{3}^{2} & = & 2\alpha_{2}\kappa_{1}\kappa_{2}\cos \theta_{2}+\frac{1}{2}\alpha_{3}\kappa_{2}^{2}+\frac{1}{2}\alpha_{1}\left(\kappa_{1}^{2}+\kappa_{2}^{2}\right)+\frac{1}{2}\rho_{3}v_{L}^{2}+\rho_{1}v_{R}^{2}\nonumber \\
 &  & +\frac{v_{L}\left[\beta_{2}\kappa_{1}^{2}\cos \theta_{L}+\beta_{1}\kappa_{2}\kappa_{1}\cos\left(\theta_{2}-\theta_{L}\right)+\beta_{3}\kappa_{2}^{2}\cos\left(2\theta_{2}-\theta_{L}\right)\right]}{2v_{R}}\,.\label{eq:LRV-57}
\end{eqnarray}
Then substituting the solutions of $\mu_{1}^{2}$ , $\mu_{2}^{2}$
, and $\mu_{3}^{2}$ into Eq.~(\ref{eq:LRV-51}) gives 
\begin{equation}
\left(v_{L}^{2}-v_{R}^{2}\right)\left[\beta_{2}\kappa_{1}^{2}\cos \theta_{L}+\cos\left(\theta_{2}-\theta_{L}\right)\beta_{1}\kappa_{2}\kappa_{1}+\cos\left(2\theta_{2}-\theta_{L}\right)\beta_{3}\kappa_{2}^{2}-\left(2\rho_{1}-\rho_{3}\right)v_{L}v_{R}\right]=0\,,\label{eq:LRV-54}
\end{equation}
which for $v_{L}^{2}-v_{R}^{2}\neq0$ leads to the seesaw relation~(\ref{eq:LRV-7}).


\end{document}